\documentstyle[aps,preprint]{revtex}

\begin{document}

\title{Theoretical Study of Time-Resolved Fluorescence
Anisotropy from Coupled Chromophore Pairs}

\author{Alexander Matro$^\dagger$ and Jeffrey A. Cina}

\address{Department of Chemistry and The James Franck Institute, The
University of Chicago, Chicago, IL 60637}

\date{}

\maketitle

\begin{center}
In press, J. Phys. Chem. (S. A. Rice issue)\\
\end{center}

\begin{abstract}
Calculations of time-resolved fluorescence anisotropy from a pair of
chromophores coupled by an excitation transfer interaction are
presented. For the purpose of investigating the effects of nuclear motion
on the energy transfer and anisotropy, an illustrative model is developed
that provides each chromophore with a single intramolecular
vibrational mode.
Account is taken of non-instantaneous excitation
and time- and frequency-resolved detection. Effects of excitation
pulse duration, detection window duration and frequency resolution, and
excitation transfer coupling strength on the time-resolved anisotropy
are examined in detail. Effects of vibrational relaxation and
\mbox{dephasing}
are also examined using a simplified Redfield description of the effects
of coupling to a thermal bath.
\end{abstract}
\vspace{20 mm}
$^\dagger$Present Address: Department of Chemistry, University of Rhode
Island, Kingston, Rhode Island 02881

\pagebreak
\section*{1. Introduction}
The process of electronic excitation transfer among chromophores
has been widely studied. \cite{silbey}
In photosynthetic antenna systems, efficient excitation transfer among
pigment molecules is responsible for greatly enhancing
the availability of energy for photosynthesis.\cite{fleming}
Studies of electronic excitation transfer among impurities in
molecular crystals have examined the effects
of the coupling between electronic and vibrational excitations.
\cite{rachsilbey}
With the availability of ultrafast lasers it is now possible to study
excitation transfer processes occurring on picosecond and
sub-picosecond timescales, enabling researchers to directly observe some of the
fastest excitation transfer processes taking place in photosynthetic
antenna systems and elsewhere.
Generally speaking, the efficiency of excitation transfer depends
on the proximity of chromophores, their spectral properties,
their relative orientations, and their interactions with the
surrounding medium.
When the chromophores coupled by excitation transfer differ in the
orientation of their transition dipole moments,
fluorescence (or stimulated emission or ground state bleaching) can
occur with polarization and frequency different from those of the
exciting field.
For this reason, measurement of some variety of
time-resolved anisotropy can often provide information on electronic
energy transport.
Two experimental methods employing linearly polarized light,
pump-probe and fluorescence upconversion,
have been used recently to obtain time-resolved anisotropies.
The time-resolved fluorescence \mbox{upconversion} signal is obtained
by combining the fluorescence induced by a laser pulse
with a delayed gate pulse in a non-linear crystal and collecting
the sum-frequency signal as a function of the delay time between
excitation and gate pulses.
By setting the polarization of the electric field of the excitation
pulse parallel or perpendicular to that of the gate pulse, parallel
($S_{\parallel}(t)$) or perpendicular ($S_{\perp}(t)$) components of
the transient fluorescence are measured.
The anisotropy is then calculated according to the expression
\begin{equation}
R(t) = \frac {S_{\parallel}(t) - S_{\perp}(t)}
{S_{\parallel}(t) + 2 S_{\perp}(t)} . \label{eq:e1.1}
\end{equation}
In pump-probe anisotropy measurements, the signal is the relative
transmittance of a delayed probe pulse.
Here the pump pulse is polarized parallel or perpendicular to a probe
pulse, which passes through the sample, rather than to a gate pulse,
which does not.

Using fluorescence upconversion,
Xie {\em et al.}\cite{xiedumets} recently measured time-resolved
anisotropy from Allophycocyanin (APC) and
C-Phycocyanin (C-PC), the pigment-protein complexes isolated from antenna
systems of photosynthetic cyanobacteria.  The structure of the trimeric
form of C-PC has been investigated with x-ray crystallography,
\cite{duerring} which
revealed that chromophores on adjacent monomers are located
within $\sim$2 nm
of one another, whereas inter-chromophore distances within individual
monomers are $\sim$5 nm.  Xie {\em et al.} observed sub-picosecond decays in
the fluorescence anisotropy in APC and C-PC.
Earlier experiments by Beck and Sauer
\cite{becksau} employing pump-probe techniques showed that
anisotropy in APC monomers fails to exhibit sub-picosecond time dependence.
Xie and co-workers attributed the ultrafast decay in trimeric APC and
C-PC to energy transfer between chromophores on adjacent monomers.
Recent femtosecond pump-probe experiments by Gillbro
{\em et al.} \cite{sharkov} have also observed a sub-picosecond
anisotropy decay in C-PC.
Rapid anisotropy decays have also been observed by Kim {\em et al.}
\cite{kim} and Zhu {\em et al.} \cite{zhugalli} in bichromophoric
compounds 9,9$^\prime$-bifluorene and
2,2$^\prime$-binaphthyl in solution using fluorescence
upconversion \cite{kim} and
pump-probe techniques. \cite{zhugalli,wynngnana}

Xie and co-workers \cite{xiedumets} made an
additional interesting observation that the
asymptotic value of the anisotropy following the ultrafast decay was
sensitive to the excitation and detection conditions used. In both C-PC and
APC, excitation at the approximate peak of the APC and C-PC absorption
spectra resulted in significantly lower long time values of the
anisotropy than excitation closer to the red edge of the
absorption spectrum, suggesting that a larger
fraction of excited chromophores transfer their excitation prior
to fluorescence in the former case.

Time-resolved anisotropy can also be a measure of coherence among
exciton states, as shown in a recent experimental study by
Galli {\em et al.} \cite{galliwynne}, where
an initial anisotropy of 0.7 was found in a pump-probe experiment on
magnesium tetraphenylporphyrin in solution.
This finding supported earlier theoretical
work by Wynne and Hochstrasser \cite{wynhoc} in which an initial anisotropy of
0.7 was predicted in a molecule having two degenerate electronic
transitions originating from a single ground state, with perpendicular
transition dipole moments.  Wynne and Hochstrasser \cite{wynhoc}
and Knox and G\"{u}len
\cite{knoxgullen} have also examined the case of two
chromophores with a fixed relative angle between their transition dipole
moments, showing that such a system can
exhibit an anisotropy larger than 0.4, the initial
anisotropy observed in a collection of randomly oriented single
chromophores.

Electronic energy transfer received a thorough theoretical
treatment by F\"{o}rster \cite{forster}, who examined the
effects of coupling strength, site energy differences, and details of potential
energy surfaces and nuclear vibrations on excitation transfer between a
pair of chromophores. Later work by Rackovsky and Silbey
focused on the effects of exciton-phonon interactions on excitation
transfer among chromophores embedded in a solid.\cite{rachsilbey}
A theoretical study by Rahman, Knox and Kenkre \cite{rahman}
investigated the fluorescence depolarization resulting from electronic
energy transfer, pointing out that electronically off-diagonal
density matrix elements need to be included in order to correctly
calculate the polarized fluorescence signal.

Previous theoretical treatments of time-resolved polarized emission from
chromophore pairs \cite{zhugalli,knoxgullen,rahman} (and single molecules
with degenerate electronic transitions \cite{wynhoc}) have neglected the
nuclear degrees of freedom, treating each chromophore as an
electronic 2-level system.
Because femtosecond pulses can launch
vibrational wave packets in both ground and excited electronic states
\cite{smithungarcina}, the effects of vibrational motion should be included
in the calculation of time-resolved signals.
In terms of the eigenstates of the uncoupled chromophore pair expressed
as products of eigenstates of individual chromophores, we can say that
Franck-Condon factors and energy level spacings between vibronic states
modify the effect of a given electronic energy transfer interaction.
\cite{forster}  In addition, different sets of states can be selected
depending on excitation and detection center frequencies and bandwidths.

In connection to the general issue of simultaneous vibrational coherence
and electronic energy transfer, it should be pointed out that
vibrational quantum beats \cite{chachi} and exciton dynamics
\cite{pullerits} have been observed in ultrafast pump-probe anisotropy
measurements on the light-harvesting complexes of {\em Rhodobacter
sphaeroides} at 4 K.  While an interesting argument was advanced
\cite{pullerits} for the observed spectral and polarization dynamics, no
attempt was reported to incorporate both nuclear and electronic dynamics
in a single model.  The light-harvesting complex LH1 of {\em Rhodobacter
sphaeroides} has also been the subject of room temperature femtosecond
fluorescence anisotropy measurements. \cite{bradforth}

Here we investigate a model chromophore pair interacting
with a simple detection apparatus
in order to explore the roles that molecular parameters and the
characteristics of the excitation and detection processes
play on the calculated time-resolved fluorescence anisotropy.
In order to study the effects of
vibrational motion on the time-resolved anisotropy, each chromophore is
considered to have a single intramolecular vibrational mode.
This model is presented in Section 2.

An analytic treatment of a chromophore pair having a minimal number of
states, presented in Section 3, highlights an important way that
chromophore vibrational structure can influence the initial anisotropy.
Recapitulating the argument of Knox and G\"{u}len
\cite{knoxgullen,rahman} it is first illustrated how interference in
emission from singly excited states with different chromophores excited
can lead to an initial anisotropy greater than 0.4.  The excited states
must, of course, emit to the same vibronic level of the electronic
ground state in order for interference to occur.  It is pointed out
that some vibronic levels are accessible from singly excited states
having only one specific chromophore excited, and that emission into
vibronic levels that reveal the identity of the excited chromophore tend
to suppress high initial values of the anisotropy.

Having included an intramolecular vibrational mode on each chromophore,
we are able to examine
the roles of simultaneous coherent vibrational and electronic
motion induced by the ultrashort excitation
laser pulse on the anisotropy. Section 4 details these numerical
calculations.  We first choose the parameters of the chromophore pair
and the excitation and detection processes to correspond as closely as
possible to the known parameters in experiments by Xie {\em et
al.}\cite{xiedumets} and compare our findings with their experimental results.
Then we investigate the effects on the anisotropy of varying the molecular and
detector
parameters, such as the strength of the excitation
transfer coupling, the difference between the zero-zero electronic transition
frequencies of two chromophores, and the duration of the
detection window.
In addition to determining time-resolved
anisotropies, we also
follow the time evolution of the excitation probabilities on the
two chromophores.
Comparing the difference in site excitation probability as a function of time
to the time-resolved anisotropy gives some indication of how
the net population transfer affects the observed anisotropy signal.

In Section 5 we examine the effects of vibronic relaxation and dephasing
resulting from the coupling of the intra-chromophore vibrations to a
thermal bath. A simplified Redfield theory is used to include these
effects.  The bath is seen to have a significant influence on
the time-resolved anisotropy, causing the anisotropy oscillations
due to coherent vibrational and excitonic motion
to diminish in amplitude.  It is found, however, that
coherent oscillations can survive the introduction of fairly rapid
vibrational relaxation. Section 6 summarizes our conclusions.

\section*{2. Model System}
We start this Section by describing the model chromophore pair.
Each chromophore has
a ground and an excited electronic state and a single vibrational degree
of freedom.  The minima of the potential energy curves for the ground
and excited electronic states are spatially displaced from each other
in the vibrational coordinate.
The Hamiltonian for chromophore 1 has the form
\begin{equation}
H_{1} = |g1\rangle H_{g1}\langle g1| + |e1\rangle (H_{e1} +
\varepsilon_{1})
\langle e1| \; . \label{eq:e2.1}
\end{equation}
where $|g1\rangle$ and $|e1\rangle$ denote the ground and
excited electronic states,
$H_{g1}$ and $H_{e1}$ denote the corresponding nuclear Hamiltonians, and
$\varepsilon_{1}$ is the zero-zero electronic transition frequency.\cite{sym}
We take $\hbar = 1$ throughout this paper and thereby express
energy in angular frequency units.
The
Hamiltonian for chromophore 2 has a form analogous to Eq. (\ref{eq:e2.1}).

The chromophore pair has four electronic states:
$|gg\rangle \: (\equiv |g1g2\rangle)$, $|eg\rangle $, $|ge\rangle $,
and $|ee\rangle $.  We omit the site index on the electronic state
except when not doing so would create ambiguity.  In $|gg\rangle $, both
chromophores are in their respective ground electronic states, $|eg\rangle $
and $|ge\rangle $ are the singly excited states with chromophore 1 and
chromophore 2 excited, respectively, and $|ee\rangle $ is the doubly excited
state.  The doubly excited state does not participate in the
electronic excitation
transfer and is omitted in what follows, but could
come into play in pump-probe measurements, where
it can be used as the final (measuring) state.\cite{wynhoc}

The two singly excited states are coupled by an excitation transfer
interaction, which in the dipole-dipole approximation would have the form
\cite{forster}
\begin{eqnarray}
V = \frac{\textstyle 1}{\textstyle R^3}
[\hat{\mbox{\boldmath $\mu$ \unboldmath}}\!_1 \cdot
\hat{\mbox{\boldmath $\mu$ \unboldmath}}\!_2
 - \frac{\textstyle 3}{\textstyle R^2}
(\hat{\mbox{\boldmath $\mu$ \unboldmath}}\!_1\cdot {\bf R})
(\hat{\mbox{\boldmath $\mu$ \unboldmath}}\!_2\cdot
 {\bf R})]
\: \cong \: U(|eg\rangle \langle ge| + h.c.) , \label{eq:e2.2}
\end{eqnarray}
where
\begin{eqnarray}
\hat{\mbox{\boldmath $\mu$ \unboldmath}}\!_1 =
\mu_1 \: {\bf e}_1 \: (|e1\rangle \langle g1|
+ h.c.) \; , \nonumber \\
\hat{\mbox{\boldmath $\mu$ \unboldmath}}\!_2 =
\mu_2 \: {\bf e}_2 \: (|e2\rangle \langle g2|
+ h.c.) \; , \label{eq:e2.moments}
\end{eqnarray}
are the electronic dipole moment operators for chromophores 1 and 2, with
$\mu_1$ and $\mu_2$ denoting the magnitudes of the electronic transition
dipole moments and ${\bf e}_1$ and ${\bf e}_2$ denoting the
unit vectors in the direction of the electronic transition dipole
moments of chromophores 1 and 2, respectively.
In the last expression of Eq. (\ref{eq:e2.2}), only the terms that
nearly conserve electronic energy have been retained, and $U$ is given
by
\begin{equation}
U = \frac{\textstyle 1}{\textstyle R^{3}_{0}} \: \mu_1 \:
\mu_2 \:[({\bf e}_1
\cdot {\bf e}_2) - \frac{\textstyle 3}{\textstyle R^{2}_{0}}
({\bf e}_1 \cdot {\bf R}_0)({\bf e}_2 \cdot {\bf R}_0)] \; . \label{eq:e2.u0}
\end{equation}
Combining Eqs. (2-5), the Hamiltonian for the
chromophore pair can be written as
\begin{equation}
H_{cp} = |gg\rangle [H_{g1}+H_{g2}]\langle gg| +
|eg\rangle [H_{e1}+H_{g2}+\varepsilon_1]\langle eg| +
|ge\rangle [H_{g1}+H_{e2}+\varepsilon_2]\langle ge| + V \; . \label{eq:e2.3}
\end{equation}
The notation we shall use to label the vibrational states of the chromophore
pair in the site
representation refers to each mode's vibrational quantum number with a
subscript designating the electronic state.  For
example, $|ge\rangle |m_{g}n_{e}\rangle $ is the state with
chromophore 2 excited, $m$ quanta of vibration in the ground state of
chromophore $1$ and $n$ quanta of vibration in the excited state of
chromophore $2$.

The interaction of the chromophore pair with the laser pulse is governed
in the dipole approximation by
\begin{equation}
V_{e}(t) = -\hat{\mbox{\boldmath $\mu$ \unboldmath}}\!_{cp}
\cdot {\bf E}(t) \; , \label{eq:e2.4}
\end{equation}
where $\hat{\mbox{\boldmath $\mu$ \unboldmath}}\!_{cp} =
\hat{\mbox{\boldmath $\mu$ \unboldmath}}\!_{1} +
\hat{\mbox{\boldmath $\mu$ \unboldmath}}\!_{2}$,
${\bf E}(t)$ is the electric field of the laser pulse,
and $\hat{\mbox{\boldmath $\mu$ \unboldmath}}\!_{1}$
and $\hat{\mbox{\boldmath $\mu$ \unboldmath}}\!_{2}$
are the electronic dipole moment operators given by
Eq. (\ref{eq:e2.moments}). We assume ${\bf E}(t)$ to be a Gaussian
laser pulse of the form
\begin{equation}
{\bf E}(t) =  E_{0}
\; {\bf e}_{L} \; \exp{(-t^{2}/2\tau_{L}^{2})}
 \cos{\Omega t} \; , \label{eq:e2.6}
\end{equation}
with amplitude $E_{0}$, width $\tau_{L}$, related to the FWHM intensity by
\begin{equation}
\tau_{L} = \frac{\textstyle FWHM}{\textstyle 2 \sqrt{ln 2}} ,
\label{eq:e2.7}
\end{equation}
center frequency $\Omega$ and polarization ${\bf e}_{L}$.

The detector consists of a collection of two-level systems with a
transition frequency distribution taken to be Gaussian.  The Hamiltonian
for the $i^{th}$ two-level detector, which is meant to play a role
analogous to that of a specific mode of the quantized radiation field,
is given by
\begin{equation}
H_d^i = |a^i\rangle E_a^i\langle a^i| +
|b^i\rangle E_b^i\langle b^i| \; ,
\label{eq:e2.8}
\end{equation}
where $|a^i\rangle $ and $|b^i\rangle $ are the ground and
excited states of the
$i^{th}$ detector two-level system, respectively, and
$\varepsilon_{d}^{i} = E_{b}^{i} - E_{a}^{i}$ is the corresponding
transition frequency.  The electronic dipole operator for the $i^{th}$
two-level system has the form
\begin{equation}
\hat{\mbox{\boldmath $\mu$ \unboldmath}}\!_{d}^i =
\mu_d \: {\bf e}_d (|a^i\rangle \langle b^i|+
|b^i\rangle \langle a^i|)  , \label{eq:e2.9}
\end{equation}
where $\mu_d$ is the magnitude of the transition dipole moment
and ${\bf e}_d$ is the unit vector specifying the detector orientation.
We omit the index $i$ on $\mu_d$ and ${\bf e}_d$, taking
all detector two-level systems to have the
same transition moment regardless of the transition frequency.  The
interaction Hamiltonian of the chromophore pair with the $i^{th}$ detector
two-level system is taken to be
\begin{equation}
V_d^i = -\eta \: \hat{\mbox{\boldmath $\mu$ \unboldmath}}\!_{cp}
\cdot \hat{\mbox{\boldmath $\mu$ \unboldmath}}\!_d^i \: ,
\label{eq:e2.10a}
\end{equation}
where $\eta$ is a constant that will cancel out in the anisotropy.
We can make the rotating wave approximation in the
chromophore pair-detector interaction, which amounts to only keeping those
terms that nearly conserve energy.  In that
approximation, Eq. (\ref{eq:e2.10a}) can be written as
\begin{eqnarray}
V_d^i=-\eta \: \mu_1 \: \mu_d \:({\bf e}_1 \cdot {\bf e}_d) \:
(|b^i\rangle |gg\rangle \langle eg|\langle a^i| + h.c.) \nonumber \\
- \eta \: \mu_2 \: \mu_d \:({\bf e}_2 \cdot {\bf e}_d) \:
(|b^i\rangle |gg\rangle \langle ge|\langle a^i| + h.c.) . \label{eq:e2.10b}
\end{eqnarray}

Time-dependent perturbation theory is used to determine the state of the
system (chromophore pair + detector) following an interaction with a
laser pulse and an interaction between the chromophore pair and
the detector. Starting in an
initial state $|gg\rangle |m_gn_g\rangle |a^i\rangle $,
in which the chromophore pair and the
detector two-level system are in their respective ground electronic
states, the probability amplitude for finding the chromophore pair in
the vibrational state $|m_g^\prime n_g^\prime\rangle $
of the ground electronic state and the detector excited to
$|b^{i}\rangle $ is given in the interaction picture by
\begin{eqnarray}
\langle b^{i}|\langle m_{g}^{\prime}n_{g}^{\prime}| \langle gg|
\widetilde{\Psi}(t)\rangle  =
- \int_{-\infty}^{t} d\tau \int_{-\infty}^{\tau} d\tau^{\prime} \nonumber \\
\times \langle b^ {i}|\langle m_{g}^{\prime}n_{g}^{\prime}|\langle gg|
\widetilde{V}_d^i(\tau) \widetilde {V}_{e}(\tau^{\prime})
 |gg\rangle |m_{g}n_{g}\rangle |a^{i}\rangle \; , \label{eq:e2.11}
\end{eqnarray}
where $\widetilde {V}_{e}(\tau)$ and $\widetilde{V}_d^i(\tau)$ are
the previously defined interaction Hamiltonians, given in
Eq. (\ref{eq:e2.4}) and Eq. (\ref{eq:e2.10b}), respectively, transformed
to the interaction picture,
\begin{equation}
\widetilde {V}_e(t) = e^{\textstyle iH_{0}t} V_e(t) e^{\textstyle -iH_{0}t}
\; , \label{eq:e2.12a1}
\end{equation}
\begin{equation}
\widetilde {V}_d^i(t) = e^{\textstyle iH_{0}t} V_d^i e^{\textstyle -iH_{0}t}
\; , \label{eq:e2.12a}
\end{equation}
and
\begin{equation}
|\widetilde {\Psi}(t)\rangle  = e^{\textstyle iH_{0}t}|\Psi(t)\rangle
\: ,\label{eq:e2.12b}
\end{equation}
with
\begin{equation}
H_{0} = H_{cp} + H^i_d . \label{eq:e2.13}
\end{equation}
Notice that the reverse ordering, in which
$\widetilde {V}_d^i(\tau^{\prime})$ acts
before $\widetilde {V}_e(\tau)$ is excluded from Eq. (\ref{eq:e2.11}) under
the rotating wave approximation of Eq. (\ref{eq:e2.10b}).
The probability of finding the chromophore pair in the electronic
ground state level $|m_{g}^{\prime}n_{g}^{\prime}\rangle $ and the detector in
$|b^{i}\rangle $ is then given by the absolute value squared of
Eq. (\ref{eq:e2.11}):
\begin{equation}
P^i(t)=|\langle b^i|\langle m_g^{\prime}n_g^\prime |\langle gg|
\widetilde {\Psi}(t)\rangle |^2 . \label{eq:e2.14}
\end{equation}
In order to calculate $P_{\parallel}^{i}(t)$ and
$P_{\perp}^{i}(t)$, we set ${\bf e}_d$ parallel and perpendicular,
respectively, to ${\bf e}_L$, and average over all possible
orientations of the chromophore pair in the laboratory frame.
For given
initial and final states of the chromophore pair, we must sum
$P^{i}(t)$ over the distribution of the detector frequencies.  Taking
that distribution to be Gaussian and converting the sum over detector
frequencies to an integral, the total excited state population of the
detector for given initial and final chromophore pair states becomes
\begin{equation}
P(t) = \int_{-\infty}^{\infty} d\varepsilon_d
D(\varepsilon_d - \overline{\varepsilon}_d) P(t,\varepsilon_d) \; ,
\label{eq:e2.15}
\end{equation}
where
\begin{equation}
D(\varepsilon_d - \overline{\varepsilon}_d) =
\frac{1}{\Delta \sqrt{\pi}}
\exp{\{-(\varepsilon_d - \overline{\varepsilon}_d)^2/\Delta^2\}} \; ,
\label{eq:e2.15D}
\end{equation}
where
$\overline{\varepsilon}_d$ is the center of of the detector frequency
distribution, $\Delta$ is its width, and $P^{i}(t)$ has been rewritten
as a function
of $\varepsilon_{d}$.

The total detector population change is obtained by
summing over all
final states of the chromophore pair and a thermal distribution
of initial states.  The time-resolved anisotropy is calculated
according to Eq. (\ref{eq:e1.1}), where the observed emission rate is obtained
from the time-dependent population of the detector using the expression
\begin{equation}
S_{\parallel(\perp)}(t) = \frac
{\textstyle P_{\parallel(\perp)}(t + \frac{\Delta t}{2})
- P_{\parallel(\perp)}(t - \frac{\Delta t}{2})}
{\textstyle \Delta t} \; , \label{eq:e2.16}
\end{equation}
in which $\Delta t$ is the response time of the detector.  By
calculating the average rate of change in detector excited state
population, as in Eq. (\ref{eq:e2.16}), we crudely incorporate the
finite time resolution of detection. \cite{melinger}

\section*{3. Simple Analytic Model}
In Section 2 we developed a model Hamiltonian with which
to calculate time-resolved anisotropy
for a pair of chromophores having multiple vibrational levels and
taking into account time and frequency resolution effects
encountered in experimental measurements.  Before proceeding to examine
these effects, we shall first investigate the properties of a far simpler
system, ignoring coherent vibrational motion by keeping only the
lowest vibrational level in each of the the singly excited electronic states,
$|eg\rangle|0_{e}0_{g}\rangle$ and
$|ge\rangle|0_{g}0_{e}\rangle$, and the three
lowest vibrational levels of the ground electronic state,
$|gg\rangle|0_{g}0_{g}\rangle$,
$|gg\rangle|1_{g}0_{g}\rangle$, and $|gg\rangle|0_{g}1_{g}\rangle$.
In their recent work, Xie\cite{xiepr}, Wynne and
Hochstrasser \cite{wynhoc}, and Knox and G\"{u}len \cite{knoxgullen}
considered similar model systems, determining that if the initial
state
(prior to excitation) and final state (following spontaneous or stimulated
emission) is $|gg\rangle|0_{g}0_{g}\rangle$, then the initial
anisotropy can take on values greater than 0.4.  The theoretical maximum
value of 0.7 is exhibited by pairs of chromophores with perpendicular
transition dipole moments of equal magnitude. Both Wynne and
Hochstrasser
\cite{wynhoc} and Knox and G\"{u}len \cite{knoxgullen} have included a
phenomenological treatment of relaxation between the excited states of the
system.
We ignore relaxation in this section, but the analytical results for the
time-resolved anisotropy of the simple model considered here will
illustrate the dependence of the anisotropy on the final state(s) of the
system.

The Hamiltonian of the simplified chromophore pair becomes
\begin{eqnarray}
H_{cp} = |gg\rangle\langle gg| \{
E_{gg}|0_{g}0_{g}\rangle\langle 0_{g}0_{g}|+
(E_{gg}+\omega)|1_{g}0_{g}\rangle\langle 1_{g}0_{g}|+
(E_{gg}+\omega)|0_{g}1_{g}\rangle\langle 0_{g}1_{g}| \}\nonumber \\
+ |eg\rangle\langle eg| E_{eg} |0_{e}0_{g}\rangle\langle 0_{e}0_{g}|
+ |ge\rangle\langle ge| E_{ge} |0_{g}0_{e}\rangle\langle 0_{g}0_{e}|
+ U(|eg\rangle\langle ge|\:|0_e0_g\rangle\langle 0_g0_e| + h.c.) \; ,
\label{eq:e3.1}
\end{eqnarray}
Diagonalizing the excited state portion of $H_{cp}$
yields the exciton states of
the system, given by \cite{cohen}
\begin{eqnarray}
|+\rangle  = \sin{\frac{\theta}{2}}|ge\rangle|0_{g}0_{e}\rangle
+ \cos{\frac{\theta}{2}}|eg\rangle|0_{e}0_{g}\rangle   \; ,
\nonumber \\
|-\rangle  = \cos{\frac{\theta}{2}}|ge\rangle|0_{g}0_{e}\rangle
- \sin{\frac{\theta}{2}}|eg\rangle|0_{e}0_{g}\rangle   \; ,
\label{eq:e3.eigen}
\end{eqnarray}
where $\tan{\theta}$ is
\begin{equation}
\tan{\theta}= \frac{2U\langle 0_{e}0_{g}|0_{g}0_{e}\rangle}
{E_{eg}-E_{ge}} , \label{eq:e2.theta}
\end{equation}
and the eigenenergies are
\begin{equation}
E_{\pm} = \frac{1}{2}(E_{eg}+E_{ge}) \pm
\frac{1}{2} \sqrt{(E_{eg}-E_{ge})^{2} + 4U^2
\langle 0_{e}0_{g}|0_{g}0_{e}\rangle^2}  \; .
\label{eq:e3.evals}
\end{equation}
The form of the dipole moment operator in the exciton representation
becomes
\begin{eqnarray}
\hat{\mbox{\boldmath $\mu$ \unboldmath}}\!_{cp} =
{\mbox{\boldmath $\mu$ \unboldmath}}\!_{+}
\{|gg\rangle|0_{g}0_{g}\rangle \langle +| + h.c.\} +
{\mbox{\boldmath $\mu$ \unboldmath}}\!_{-}
\{|gg\rangle|0_{g}0_{g}\rangle \langle -| + h.c.\} \nonumber \\
+ {\mbox{\boldmath $\mu$ \unboldmath}}\!_{1}
\langle 1_g0_g|0_e0_g\rangle
\{\cos{\frac{\theta}{2}} |gg\rangle|1_g0_g\rangle\langle +| -
\sin{\frac{\theta}{2}} |gg\rangle|1_g0_g\rangle\langle -| + h.c. \}
\nonumber \\
+ {\mbox{\boldmath $\mu$ \unboldmath}}\!_{2}
\langle 0_g1_g|0_g0_e\rangle
\{\sin{\frac{\theta}{2}} |gg\rangle|0_g1_g\rangle\langle +| +
\cos{\frac{\theta}{2}} |gg\rangle|0_g1_g\rangle\langle -| + h.c. \}
 \; , \label{eq:e3.mucp}
\end{eqnarray}
with ${\mbox{\boldmath $\mu$ \unboldmath}}\!_{+}$
and
${\mbox{\boldmath $\mu$ \unboldmath}}\!_{-}$
given by
\begin{eqnarray}
{\mbox{\boldmath $\mu$ \unboldmath}}\!_{+} & = &
{\mbox{\boldmath $\mu$ \unboldmath}}\!_{2}
 \langle 0_g0_g|0_g0_e \rangle \sin{\frac{\theta}{2}} +
{\mbox{\boldmath $\mu$ \unboldmath}}\!_{1}
 \langle 0_g0_g|0_e0_g \rangle \cos{\frac{\theta}{2}}, \nonumber \\
{\mbox{\boldmath $\mu$ \unboldmath}}\!_{-} & = &
{\mbox{\boldmath $\mu$ \unboldmath}}\!_{2}
 \langle 0_g0_g|0_g0_e \rangle \cos{\frac{\theta}{2}} -
{\mbox{\boldmath $\mu$ \unboldmath}}\!_{1}
 \langle 0_g0_g|0_e0_g \rangle \sin{\frac{\theta}{2}} \; .
\label{eq:e3.muplmin}
\end{eqnarray}

In addition to simplifying the chromophore pair model, we also simplify
the interaction with the laser pulse and the detector by assuming that
the excitation pulse duration and the detection window are both much shorter
than the timescales of dynamics in the excited electronic state.
Under these conditions,
the rate of emission, $S(t)$,
can be obtained by differentiating both sides of Eq. (\ref{eq:e2.15})
with respect to time,
\begin{equation}
S(t)=
\int_{-\infty}^{\infty} d\varepsilon_d
D(\varepsilon_d - \overline{\varepsilon}_d)
S(t,\varepsilon_d) \: , \label{eq:e3.4}
\end{equation}
where $S(t,\varepsilon_d)$ is the time derivative of $P(t,\varepsilon_d)$
given by Eq. (\ref{eq:e2.14}) and
$D(\varepsilon_d - \overline{\varepsilon}_d)$ is defined in
Eq. (\ref{eq:e2.15D}).
The probability amplitude
for finding the
system in $|gg\rangle|0_{g}0_{g}\rangle|b\rangle$, starting in the
initial state $|gg\rangle|0_{g}0_{g}\rangle|a\rangle$, is given
in the interaction picture by
\begin{eqnarray}
\langle b|\langle0_{g}0_{g}|\langle gg|\widetilde{\Psi}(t)\rangle =
- \frac{1}{\sqrt{2}}\sqrt{\pi}\tau_L E_0 \eta \mu_+^2 \mu_d
({\bf e}_+ \cdot {\bf e}_{L})
({\bf e}_+ \cdot {\bf e}_d)
\int_{0}^{t}d\tau \exp{i(\varepsilon_d - \varepsilon_{+g})\tau} \nonumber \\
- \frac{1}{\sqrt{2}}\sqrt{\pi}\tau_L E_0 \eta \mu_-^2 \mu_d
({\bf e}_- \cdot {\bf e}_{L})
({\bf e}_- \cdot {\bf e}_d)
\int_{0}^{t}d\tau \exp{i(\varepsilon_d - \varepsilon_{-g})\tau} \: ,
\label{eq:e3.5}
\end{eqnarray}
where $\varepsilon_{\pm g} = E_\pm - E_{gg}$ and
$\mbox{\boldmath $\mu$ \unboldmath}\!_{\pm} \equiv \mu_\pm {\bf e}_\pm$.
Calculating $S(t,\varepsilon_d)$, integrating over
the frequency profile of the detector according to Eq. (\ref{eq:e3.4}),
and taking $\Delta \gg |\varepsilon_{+-}|$ where
$\varepsilon_{+-} = E_+ - E_-$ yields
\begin{eqnarray}
S(t) = \frac{1}{2}\pi\tau_L^2 E_0^2 \eta^2 \mu_+^4 \mu_d^2
({\bf e}_+ \cdot {\bf e}_{L})^{2}({\bf e}_+ \cdot {\bf e}_d)^{2} +
\frac{1}{2}\pi\tau_L^2 E_0^2 \eta^2 \mu_-^4 \mu_d^2
({\bf e}_- \cdot {\bf e}_{L})^{2}({\bf e}_- \cdot {\bf e}_d)^{2} \nonumber \\
+ \: {\pi\tau_L^2 E_0^2 \eta^2 \mu_+^2\mu_-^2 \mu_d^2}
({\bf e}_+ \cdot {\bf e}_{L})({\bf e}_+ \cdot {\bf e}_d)
({\bf e}_- \cdot {\bf e}_{L})({\bf e}_- \cdot {\bf e}_d)
\cos{(\varepsilon_{+-}t)} \: . \label{eq:e3.7}
\end{eqnarray}
The first two terms in Eq. (\ref{eq:e3.7}) result from excitation and
emission via $|+\rangle$ and $|-\rangle$, respectively,
and the third (oscillatory)
term results from interference in emission from
$|+\rangle$ and $|-\rangle$.
While the internal geometry of the chromophore pair is assumed to
be fixed, it can be randomly oriented in space.  To average Eq.
(\ref{eq:e3.7}) over spatial orientations, we integrate ${\bf e}_+$
over $4\pi$ of solid angle and, maintaining a fixed angle between
${\bf e}_+$ and ${\bf e}_-$, integrate the latter over $2\pi$ radians.
Defining a direction cosine
\begin{equation}
\cos{\gamma} = {\bf e}_+ \cdot {\bf e}_- \: ,
\label{eq:cosgamma}
\end{equation}
the orientationally averaged signal is given by (see Appendix)
\begin{equation}
S_{\parallel}^{avg}(t) = \frac{\pi\tau_L^2 E_0^2 \eta^2 \mu_d^2}{2} \{
\frac{1}{5}(\mu_+^4 + \mu_-^4)
+2\mu_+^2\mu_-^2(\frac{1}{5} {\cos}^2 \gamma +
\frac{1}{15} {\sin}^2 \gamma)\cos{(\varepsilon_{+-}t)}\} \; ,
\label{eq:e3.savgpar}
\end{equation}
\begin{equation}
S_{\perp}^{avg}(t) = \frac{\pi\tau_L^2 E_0^2 \eta^2 \mu_d^2}{2} \{
\frac{1}{15}(\mu_{+}^4 + \mu_-^4)
+2\mu_+^2\mu_-^2(\frac{1}{15} {\cos}^2 \gamma -
\frac{1}{30} {\sin}^2 \gamma)\cos{(\varepsilon_{+-}t)}\} \: .
\label{eq:e3.savgperp}
\end{equation}
Using Eqs. (\ref{eq:e3.savgpar}) and (\ref{eq:e3.savgperp})
in Eq. ({\ref{eq:e1.1}) gives the time-dependent anisotropy
\begin{equation}
R(t) = \frac
{
2(\mu_+^4+\mu_-^4) + \mu_+^2\mu_-^2(4+\cos^2 \gamma)
\cos{(\varepsilon_{+-} t)}
}
{
5(\mu_+^4+\mu_-^4) + 10 \mu_+^2\mu_-^2\cos^2 \gamma
\cos{(\varepsilon_{+-} t)}
} \; . \label{eq:e3.roftgen}
\end{equation}

Let us now examine the case of identical chromophores.
In that case, and assuming $\theta = \frac{\textstyle\pi}{\textstyle 2}$
in Eq. (\ref{eq:e2.theta}), the
expressions for $\mu_{+}$ and $\mu_{-}$ in Eq. (\ref{eq:e3.muplmin})
simplify to
\begin{equation}
\mu_{\pm}{\bf e}_\pm = \frac{1}{\sqrt{2}}\langle 0_g0_g|0_e0_g \rangle
(\mu_{2}{\bf e}_2 \pm \mu_{1}{\bf e}_1) \: . \label{eq:e3.mupmperp}
\end{equation}
Because $\mu_1=\mu_2$,
the two transition dipole moments in Eq. (\ref{eq:e3.mupmperp})
are perpendicular.
Substituting
$\gamma= \frac{\textstyle \pi}{\textstyle 2}$ into
Eq. (\ref{eq:e3.roftgen}) yields
\begin{equation}
R(t) = \frac{2(\mu_+^4 + \mu_-^4)
\:+\: 3\,
\mu_+^2\mu_-^2 \cos{(\varepsilon_{+-}t)}}
{5(\mu_+^4 + \mu_-^4)} \: . \label{eq:e3.rsimple}
\end{equation}
The expression (\ref{eq:e3.rsimple}) for the time-resolved anisotropy
can be restated in terms of the
angle between the directions of the transition dipole moments of the
two chromophores, ${\bf e}_1$ and ${\bf e}_2$. Calling
that angle $\phi$, Eq. (\ref{eq:e3.rsimple}) becomes (see Appendix)
\begin{equation}
R(t) = 0.4 \:+ \: 0.3
\frac{1-\cos^2\phi}{1+\cos^2\phi}\cos{(\varepsilon_{+-}t)}
\; . \label{eq:e3.8}
\end{equation}

The denominator of Eq. (\ref{eq:e3.rsimple}) is proportional to the
total intensity and is time-independent.  The time-independent overall
intensity
results from the fact that the chromophore pairs are identical and randomly
distributed.
When transitions to both exciton states are optically bright, the
oscillatory term initially enhances the parallel signal while
diminishing the
perpendicular signal, resulting in an initial anisotropy larger than $0.4$.
Alternatively, if one of the exciton states is optically dark, the
interference term vanishes and the anisotropy is $0.4$ and
time-independent.  The maximum anisotropy can be attained when the
interference term is maximized, which occurs when ${\bf e}_1$ and
${\bf e}_2$ are perpendicular.

Let us now compare the anisotropy in Eq. (\ref{eq:e3.8}) with the
anisotropy that would be obtained if the only admitted final states of the
chromophore pair were $|gg\rangle|1_{g}0_{g}\rangle$ and
$|gg\rangle|0_{g}1_{g}\rangle$, the two electronic ground
states with one of the chromophores in the first vibrationally excited
state.  The probability amplitude of finding the system in
$|gg\rangle|1_{g}0_{g}\rangle|b\rangle$,
starting in the initial state $|gg\rangle|0_{g}0_{g}\rangle|a\rangle$
with an excitation pulse short on the timescale of
excited state dynamics, is
\begin{eqnarray}
\langle b|\langle1_g0_g|\langle gg|\widetilde{\Psi}(t)\rangle =
- \frac{1}{\sqrt{2}}\sqrt{\pi}\tau_L E_0\eta\mu_d \cos \frac{\theta}{2}
\langle0_e0_g|1_g0_g\rangle \mu_+\mu_1
({\bf e}_+ \cdot {\bf e}_L)({\bf e}_1 \cdot {\bf e}_d) \nonumber \\
\times\int_{0}^{t}d\tau \exp{i(\varepsilon_d - \varepsilon_{+g} + \omega)\tau}
\nonumber \\
+ \frac{1}{\sqrt{2}}\sqrt{\pi}\tau_L E_0\eta\mu_d \sin \frac{\theta}{2}
\langle0_e0_g|1_g0_g\rangle \mu_-\mu_1
({\bf e}_- \cdot {\bf e}_L)({\bf e}_1 \cdot {\bf e}_d) \nonumber \\
\times\int_{0}^{t}d\tau \exp{i(\varepsilon_d - \varepsilon_{-g} + \omega)\tau}
\:. \label{eq:e3st1}
\end{eqnarray}
Since the
Franck-Condon factor between states $|0_{g}0_{e}\rangle$ and
$|1_{g}0_{g}\rangle$
vanishes, there is no ${\bf e}_2$ component present in emission from
$|+\rangle$ and $|-\rangle$ into $|gg\rangle|1_{g}0_{g}\rangle$.
The rate of emission, obtained in the same manner as Eq.
(\ref{eq:e3.7}), is
\begin{eqnarray}
S(t) = \frac{1}{2}\pi\tau_L^2E_0^2 \eta^2 \mu_+^2 \mu_1^2 \mu_d^2 \cos^2
\frac{\theta}{2}
\langle0_e0_g|1_g0_g\rangle^2 \: \overline{
({\bf e}_+ \cdot {\bf e}_L)^2({\bf e}_1 \cdot {\bf e}_d)^2}
\nonumber \\
+ \frac{1}{2} \pi\tau_L^2 E_0^2 \eta^2 \mu_-^2 \mu_1^2 \mu_d^2 \sin^2
\frac{\theta}{2}
\langle0_e0_g|1_g0_g\rangle^2 \: \overline{
({\bf e}_- \cdot {\bf e}_L)^2({\bf e}_1 \cdot {\bf e}_d)^2} \nonumber \\
+ \pi\tau_L^2 E_0^2 \eta^2 \mu_+ \mu_- \mu_1^2 \mu_d^2
\sin \frac{\theta}{2} \cos \frac{\theta}{2}
\langle0_e0_g|1_g0_g\rangle^2 \:
\overline{({\bf e}_+ \cdot {\bf e}_L)
({\bf e}_- \cdot {\bf e}_L)({\bf e}_1 \cdot {\bf e}_d)^2}
\cos{(\varepsilon_{+-}t)} \; , \label{eq:e3.1g0grate}
\end{eqnarray}
where the overbar signifies orientational averaging.
The rate of emission into $|gg\rangle|0_{g}1_{g}\rangle$ has an
identical form to Eq. (\ref{eq:e3.1g0grate}).
The time-resolved
anisotropy for the case of identical chromophores now takes the form
\begin{equation}
R(t) = \frac
{2(\mu_+^4 + \mu_-^4) + \mu_+^2 \mu_-^2(6\cos{(\varepsilon_{+-}t)} - 2)}
{5(\mu_+^2 + \mu_-^2)^2} \label{eq:e3.9a}
\end{equation}
in terms of the transition moments of the exciton states or,
alternatively,
\begin{equation}
R(t) = 0.1 \: + \: 0.3\cos^2\phi \: + \:
0.3(1-\cos^2\phi)\cos{(\varepsilon_{+-}t)}
 \: , \label{eq:e3.9}
\end{equation}
where $\phi$ is the angle between ${\bf e}_1$ and ${\bf e}_2$.
The time zero anisotropy predicted by Equations
(\ref{eq:e3.9a}) and
(\ref{eq:e3.9}) is $0.4$, independent of the angle
between ${\bf e}_1$ and ${\bf e}_2$.

The qualitative difference between anisotropies described by Eqs.
(\ref{eq:e3.8}) and (\ref{eq:e3.9}) at \mbox{$t=0$} can be understood by
considering Eqs. (\ref{eq:e3.7}) and (\ref{eq:e3.1g0grate}) in the
short time limit where $\cos(\varepsilon_{+-}\delta t) \cong 1$.
The emission
rate in Eq. (\ref{eq:e3.7}) at $\delta t$ becomes
\begin{eqnarray}
S(\delta t) \cong
\frac{1}{2} \pi\tau_L^2 E_0^2\eta^2\mu_1^4\mu_d^2
\langle 0_g0_g|0_e0_g\rangle^2 \: \overline{
({\bf e}_1\cdot{\bf e}_L)^2({\bf e}_1\cdot{\bf e}_d)^2}
\nonumber \\
+\frac{1}{2} \pi\tau_L^2 E_0^2\eta^2\mu_2^4\mu_d^2
\langle 0_g0_g|0_g0_e\rangle^2 \: \overline{
({\bf e}_2\cdot{\bf e}_L)^2({\bf e}_2\cdot{\bf e}_d)^2}
\nonumber \\
+ \pi\tau_L^2 E_0^2\eta^2\mu_1^2\mu_2^2\mu_d^2
\langle 0_g0_g|0_e0_g\rangle
\langle 0_g0_g|0_g0_e\rangle \: \overline{
({\bf e}_1\cdot{\bf e}_L)({\bf e}_2\cdot{\bf e}_L)
({\bf e}_1\cdot{\bf e}_d)({\bf e}_2\cdot{\bf e}_d)} \; ,
\label{eq:e3.sh1}
\end{eqnarray}
where we have used Eq. (\ref{eq:e3.muplmin}) to express the rate in
terms of the site transition dipole moments.  Carrying out the
orientational average (see Eq. (\ref{eq:ea6}) of the Appendix) and calculating
the
the anisotropy yields
\begin{equation}
R(\delta t) = \frac{2\mu_1^4 \langle 0_g0_g|0_e0_g\rangle^2
+ 2\mu_2^4 \langle 0_g0_g|0_g0_e\rangle^2
+ \mu_1^2\mu_2^2
(\cos^2 \phi + 3)\langle 0_g0_g|0_e0_g\rangle\langle
0_g0_g|0_g0_e\rangle}
{5\mu_1^4 \langle 0_g0_g|0_e0_g\rangle^2
+ 5\mu_2^4 \langle 0_g0_g|0_g0_e\rangle^2
+ 10\mu_1^2\mu_2^2
\cos^2 \phi \langle 0_g0_g|0_e0_g\rangle\langle
0_g0_g|0_g0_e\rangle} \;, \label{eq:e3.rshort1}
\end{equation}
where $\phi$ is the angle between ${\bf e}_1$ and ${\bf e}_2$.
The first two terms in the numerator and the denominator of
Eq. (\ref{eq:e3.rshort1}) arise from contributions involving each
chromophore separately, while the third term results from contributions
from both chromophores.  This crossterm depends on the angle between
the dipole moments of the chromophores, and the short-time anisotropy has its
maximum when $\phi = \frac{\textstyle \pi}{\textstyle 2}$.
On the other hand, the rate of emission into $|gg\rangle|1_g0_g\rangle$,
given in Eq. (\ref{eq:e3.1g0grate}), becomes
\begin{equation}
S(\delta t) \cong
\frac{1}{2} \pi\tau_L^2 E_0^2\eta^2\mu_1^4\mu_d^2
\langle 0_g0_g|0_e0_g\rangle\langle 0_e0_g|1_g0_g\rangle \: \overline{
({\bf e}_1\cdot{\bf e}_L)^2({\bf e}_1\cdot{\bf e}_d)^2} \; ,
\label{eq:e3.sh2}
\end{equation}
where Eq. (\ref{eq:e3.muplmin}) was again used to express the rate
in terms of transition dipole moments of the individual monomers.
Carrying out the orientational average and calculating the anisotropy
yields $R(\delta t) = 0.4$.
There is no crossterm in Eq. (\ref{eq:e3.sh2})
analogous to the one in Eq. (\ref{eq:e3.sh1}) because transitions from
$|eg\rangle|0_{e}0_{g}\rangle$  to
$|gg\rangle|0_{g}1_{g}\rangle$ and from $|ge\rangle|0_{g}0_{e}\rangle$
to $|gg\rangle|1_{g}0_{g}\rangle$ are Franck-Condon forbidden.
There is only one route to each final state in this case, and quantum
mechanical interference in emission from the two sites does not occur.
Therefore, the $t=0$ anisotropy given by Eq. (\ref{eq:e3.8}) is sensitive to
the presence of both chromophores, while the anisotropy given in Eq.
(\ref{eq:e3.9}) is not.
It is worth emphasizing that the high initial value of the anisotropy
predicted by Eq. (\ref{eq:e3.rshort1}) reflects the fixed relative
geometry of the chromophore pair, rather than the presence of energy
transfer, {\em per se}.  The lack of dependence of $S(\delta t)$ in Eqs.
(\ref{eq:e3.sh1}) and (\ref{eq:e3.sh2}) on the exciton splitting is in
keeping with that fact.

Although it might appear in Eq. (\ref{eq:e3.8}) as though $R(t)$ would
exhibit oscillations whenever $\phi \neq 0$, this is not always the case.
For example, in the case when $\phi = \frac{\textstyle\pi}{\textstyle 2}$
and either ${\bf e}_1$ or
${\bf e}_2$ lies along the inter-chromophore axis, the energy splitting between
the exciton states will be zero, assuming
dipolar coupling of the form given in Eq. (\ref{eq:e2.2}),
and the anisotropy will be
time-independent.  Conversely, excitation transfer is not always
required for the anisotropy to exhibit oscillations.
The anisotropy
given by Eq. (\ref{eq:e3.roftgen}), which applies to chromophores that
are not necessarily identical,
will exhibit oscillations even in the case
of zero excitation transfer coupling, as long as the
individual chromophores have different electronic transition
frequencies.

 From the calculation presented in this Section we see that time-resolved
anisotropy in a \mbox{multilevel} system will contain both kinds of
contributions described by Eqs. (\ref{eq:e3.8}) and (\ref{eq:e3.9}), and
that
proper treatment of time-resolved fluorescence anisotropy must include
the multilevel structure of both the excited and ground electronic
states of the chromophores.
The relative weights of the contributions leading to the anisotropies
given in Eqs. (\ref{eq:e3.8}) and (\ref{eq:e3.9}) will be determined by
the spectral characteristics of the excitation and detection processes.
Generally speaking, contributions giving
rise to the standard ($0.4$) value of the initial anisotropy are
expected to outweigh the contributions of the kind described by Eq.
(\ref{eq:e3.8}), especially when higher vibrational states of the ground
electronic state are selected as the final states by the detection
process, which is often the case with fluorescence \mbox{upconversion}.
\cite{xiedumets}

\section*{4. Effects of Vibrational Motion and Excitation/Detection
Conditions}
The analysis of Section 3, despite using a simplified model
chromophore pair and neglecting effects of
less-than-ideal excitation and detection
conditions, serves to
demonstrate the importance of properly including the
vibrational structure of
initial and final states when calculating the fluorescence anisotropy.
In this Section, we will utilize the model developed in Section 2 more
fully, including the full vibrational level structure of both chromophores and
non-zero duration for the excitation pulse and the detection window.

The model in Section 2 was developed in the site representation in order
to give a better physical picture of different features of the model.
The calculations in this Section, like those in Section 3,
are more conveniently
carried out in the energy eigenstate representation.  The chromophore pair
Hamiltonian in the eigenstate representation can be written as
\begin{equation}
H_{cp} = \sum_{\alpha} |\alpha\rangle E_{\alpha}\langle\alpha| +
\sum_{\gamma} |\gamma\rangle E_\gamma\langle\gamma| \label{eq:e4.10}
\end{equation}
where, for simplicity of notation, we denote all eigenstates in the ground
electronic \mbox{manifold} of the chromophore pair by $\alpha$ and all exciton
states by $\gamma$, with $E_{\alpha}$ and $E_{\gamma}$ as the
corresponding angular frequencies.  The electronic dipole
operator can be expressed as
\begin{equation}
\hat{\mbox{\boldmath $\mu$ \unboldmath}}\!_{cp} =
\sum_{\alpha , \gamma} \mu_{\alpha \gamma}
{\bf e}_{\alpha \gamma}
(|\alpha\rangle\langle\gamma| + |\gamma\rangle\langle\alpha|) \; ,
\label{eq:e4.11}
\end{equation}
where $\mu_{\alpha \gamma}{\bf e}_{\alpha \gamma}$ is the
transition dipole moment between
states $|\alpha\rangle$ and $|\gamma\rangle$.

Starting in the initial state $|\alpha\rangle|a\rangle$, the probability
amplitude for finding the system in a final state
$|\alpha^{\prime}\rangle|b\rangle$ following an interaction
with a laser pulse and
the detector is given by (see Eq. (\ref{eq:e2.11}))
\begin{eqnarray}
\langle b|\langle \alpha^\prime|\widetilde{\Psi}(t)\rangle
= - \frac{E_0\eta \mu_d}{2}\sum_{\gamma}
\mu_{\alpha \gamma}\mu_{\alpha^{\prime} \gamma}
({\bf e}_{\alpha \gamma}\cdot {\bf e}_L)
({\bf e}_{\alpha^{\prime} \gamma} \cdot {\bf e}_d) \int_{-\infty}^{t} d\tau
\int_{-\infty}^{\tau} d\tau_{1} \nonumber \\
\times\exp{\{i(\varepsilon_d-\varepsilon_{\gamma \alpha^\prime})\tau\}}
\exp{\{i(\varepsilon_{\gamma \alpha}-\Omega)\tau_1\}}
\exp{\{-\tau_1^2/2\tau_L^2\}} \; . \label{eq:e4.12}
\end{eqnarray}
Calculating the
excited state population in $|b\rangle$ according to Eq. (\ref{eq:e2.14})
and integrating over the frequency profile of the detector according to
Eq. (\ref{eq:e2.15}) yields
\begin{eqnarray}
P(t,\overline{\varepsilon}_d) =
\frac{E_0^2\eta^2 \mu_d^2}{4}
\sum_{\gamma , \gamma^{\prime}}
\mu_{\alpha \gamma}\mu_{\alpha \gamma^{\prime}}
\mu_{\alpha^{\prime} \gamma}
\mu_{\alpha^{\prime} \gamma^{\prime}}
({\bf e}_{\alpha \gamma}\cdot {\bf e}_L)
({\bf e}_{\alpha \gamma^{\prime}}\cdot {\bf e}_L)
({\bf e}_{\alpha^{\prime} \gamma} \cdot {\bf e}_d)
({\bf e}_{\alpha^{\prime} \gamma^{\prime}} \cdot {\bf e}_d) \nonumber \\
\times\int_{-\infty}^{t} d\tau \int_{-\infty}^{\tau} d\tau_1
\int_{-\infty}^{t} d\tau_2 \int_{-\infty}^{\tau_{2}} d\tau_3
\exp{\{-(\tau-\tau_2)^2\Delta^2/4\}} \nonumber \\
\times\exp{\{-i(\overline{\varepsilon}_d -
\varepsilon_{\gamma^\prime \alpha^\prime})\tau\}}
\exp{\{-i(\varepsilon_{\gamma^\prime \alpha}-\Omega)\tau_1\}}
\exp{\{-\tau_1^2/2\tau_L^2\}} \nonumber \\
\times\exp{\{i(\overline{\varepsilon}_d -
\varepsilon_{\gamma \alpha^\prime} )\tau_2\}}
\exp{\{i(\varepsilon_{\gamma \alpha}-\Omega)\tau_3\}}
\exp{\{-\tau_3^2/2\tau_L^2\}} \label{eq:e4.13}
\end{eqnarray}
As discussed in Section 2, the total excited state population is
calculated by summing Eq. (\ref{eq:e4.13}) over all final states and
thermally weighted initial states of the chromophore pair.

The average, over chromophore pair orientations, of the products of four
direction cosines appearing in Eq. (\ref{eq:e4.13}) must now be carried
out.
It is convenient to express the
transition moments ${\bf e}_{\alpha \gamma}\mu_{\alpha \gamma}$
in terms of ${\bf e}_1\mu_{1}$ and ${\bf e}_2\mu_{2}$,
the transition dipole moments of the individual chromophores,
\begin{equation}
{\bf e}_{\alpha \gamma}\mu_{\alpha \gamma} =
\langle \alpha |gg\rangle\langle eg|\gamma\rangle {\bf e}_1\mu_{1}
+ \langle \alpha |gg\rangle\langle ge|\gamma\rangle {\bf e}_2\mu_{2}
\; . \label{eq:e4.14}
\end{equation}
Using Eq. (\ref{eq:e4.14}) enables us to express the product of four
direction cosines in Eq. (\ref{eq:e4.13}) in terms of direction cosines
involving ${\bf e}_1$ and ${\bf e}_2$. There are four different
orientational averages to perform. These averages are given in the
Appendix. Eq. (\ref{eq:e4.13}) can now be used to compute the population
in the excited state of the detector, and subsequently the
time-resolved fluorescence anisotropy using Eq. (\ref{eq:e2.16}).

In the numerical calculation presented in this Section, we have chosen
the parameters for the chromophore pair to roughly correspond to the
known parameters in the closely-coupled chromophore pairs consisting of
$\alpha$-84 and $\beta$-84 tetrapyrroles on adjacent monomers in C-PC
trimers.\cite{xiedumets}
The value of the excitation transfer coupling matrix
element is set to -50 cm$^{-1}$ (1 cm$^{-1}$ corresponds to an angular
frequency of $1.885\times 10^{11} s^{-1}$),
corresponding to
Sauer and Scheer's calculation of the excitonic interaction
energy.\cite{saucheer} The angle between the transition moments of the
chromophores is set to $65^{o}$, corresponding to the value estimated
from the x-ray crystallographic structure. \cite{duerring}
The zero-zero electronic transition
frequency of chromophore 2 is taken to be 150 cm$^{-1}$ higher than that of
chromophore 1.  This offset corresponds to the difference in absorption
maxima of the $\alpha$-84 and $\beta$-84 chromophores.\cite{saucheer,sss}
The potential energy curves for $H_{g1}$,
$H_{g2}$, $H_{e1}$, and $H_{e2}$ are chosen to be harmonic and to have
the same frequency of 100 cm$^{-1}$.  This choice of vibrational
frequency is arbitrary, but is typical of a low frequency molecular
vibration and is in the interesting range comparable to the frequencies of
pure exciton dynamics.
In the absence of experimental data, we assume a
small displacement between ground and excited potential energy curves on
each chromophore of 1.5 times the rms width of the ground state
coordinate distribution, $\frac{1}{\sqrt{2\omega}}$ for a
mass-weighted oscillator coordinate.
The excitation laser pulse is taken to be 70 fs Full
Width at Half Maximum (FWHM) in intensity ($\tau_{L}=42 \: fs$), in
accordance with experiment.\cite{xiedumets}  The frequency resolution of
the detector is set to 210 cm$^{-1}$ FWHM in intensity ($\Delta$ = 126
cm$^{-1}$), which corresponds to the frequency width of the
detection window used in the experiments,\cite{xiedumets}
and the width of the time detection window is taken to be 70 fs.
The radiative
lifetimes of the excited states are much longer than our timescale of
interest of a few picoseconds, and non-radiative excited state
population decay mechanisms, such as intersystem crossing, are ignored.
The initial state of the chromophore pair is
$|gg\rangle |0_g0_g\rangle$ for all calculations in this section.  See
Section 5 for calculations that include additional thermally populated
initial states.

Figure 1 shows time-dependent anisotropies calculated for the model
system using the parameters listed above.
Two different sets of excitation and detection conditions are
illustrated.  The upper (thick) curve in Figure 1 was calculated
with $\Omega$, the center frequency of the laser pulse, resonant with
$\varepsilon_1$, the zero-zero electronic transition frequency of
chromophore 1.  The lower (thin) curve was calculated
with $\Omega$ at 100 cm$^{-1}$ above $\varepsilon_2$, the zero-zero
electronic transition frequency of the chromophore 2.
In each case, the average detector frequency,
$\overline{\varepsilon}_d$, was centered 250 cm$^{-1}$ below $\Omega$.
We included 30 states in the singly excited manifold, allowing up to 4
quanta of combined vibrational excitation,
and 28 states in the ground manifold,
allowing up to 6 quanta of combined vibrational excitation.  Inclusion of
additional states in the ground or the excited manifold had no effect on
the calculated anisotropy.

The difference in excitation and detection conditions in the two cases
gives rise to qualitative differences in the anisotropy.
Anisotropies in Figure 1 exhibit complicated behavior that
contains effects of both exciton dynamics (including excitation transfer)
and nuclear dynamics.
It can be seen from the near constancy of
the anisotropy that little change occurs in the emission
polarization direction when
the excitation pulse only weakly populates the excited states of the
higher-energy chromophore (thick curve).
The absence of exciton dynamics is reflected
in the weak time-dependence and relatively high value of the anisotropy.
On the other hand, anisotropy resulting from excitation at 100 cm$^{-1}$
above the zero-zero electronic
transition frequency of the higher-energy chromophore, shown by the
lower (thin) curve in Figure 1, resonantly populates a range of states
having amplitude on both
chromophores, leading to a larger extent of excitation transfer, as
evidenced by the large oscillations in the anisotropy.  The dependence
of the anisotropy on the excitation frequency in our calculations shows
a similar trend to the experimental results of Xie {\em et al.}
\cite{xiedumets}, who found that excitation
at the red edge of the absorption spectrum of the chromophore pairs in
APC and C-PC trimers resulted in a more weakly time-dependent anisotropy
with higher asymptotic values than excitation at a higher frequency.

One feature common to the two plots in Figure 1 is their high
initial anisotropy and subsequent rapid decrease during the first 0.1
ps.  Despite the difference in site energy between the chromophores and
the finite bandwidth of the excitation pulse and the detector, both
chromophores can be excited at short times and the interference in
emission necessary for anisotropy greater than 0.4 can occur.  When the
excitation pulse and the detection window overlap, both are effectively
shortened by the requirement that the former must act before the
latter.  In addition, the nuclear wave packets $|\psi_{eg}(t)\rangle$
and $|\psi_{ge}(t)\rangle$ prepared in the Franck-Condon regions of the
two excited state potential energy surfaces both resemble
$|0_g0_g\rangle$ at short times, favoring emission into the same ground
state.  Thus, the initial anisotropy is dominated by terms of the kind
illustrated by Eq. (\ref{eq:e3.8}).

The effective frequency resolution improves once the pulses cease to
overlap and each can operate for its full duration.  Moreover, since
$|\psi_{eg}(t)\rangle$ and $|\psi_{ge}(t)\rangle$ move differently on
their respective potential energy surfaces, nuclear motion favors
emission to non-overlapping final states, decreasing the effect of
quantum mechanical interference in the anisotropy.  It should also be
mentioned that the site energy difference between the two chromophores
contributes to a differing rate of phase accumulation, which leads to to
an overall cosinusoidal oscillation in the anisotropy.
All of these factors
can contribute to a rapid decrease in the anisotropy, even in the
absence of energy transfer coupling between the sites.  For this
reason, the rapid ``decay'' of the anisotropies of Figure 1 cannot
automatically be regarded as a manifestation of ultrafast energy transfer.

In the calculations presented in Figure 1, the time detection window
$\Delta t$ was taken to be 70 fs, equal to the FWHM duration of
the excitation pulse and the time corresponding to the FWHM frequency
width of the detector.  However, experimental time resolution can often
be lower than the inverse frequency width of the detector.  Figure 2
compares the anisotropy shown in the lower (thin) curve of Figure 1 with
the anisotropy calculated using identical parameters except that the time
detection window, $\Delta t$, was lengthened to 150 fs.  The $t=0$
anisotropy drops considerably as the time resolution decreases.
Therefore, limitations in the experimental time resolution can prevent
observation of the high initial anisotropy discussed above.

Figure 3 shows time-dependent anisotropies calculated with the
same parameters as those used in Figure 1 except that the
offset between the zero-zero electronic transition frequencies of the two
chromophores is decreased to 100 cm$^{-1}$ (i.e. equal to the
vibrational level spacing) by decreasing the zero-zero transition frequency
$\varepsilon_2$ of the higher-energy chromophore, resulting in more efficient
excitation transfer because many vibronic levels of the two chromophores are
now
in resonance.  Because the excitation frequencies were kept the same as
those in Figure 1, the $\Omega$ value for the lower (thin) curve in Figure 3 is
now 150 cm$^{-1}$ above the zero-zero electronic transition frequency of
the higher-energy chromophore.
The time-resolved anisotropies in Figure 3 show
considerable differences from those of Figure 1.  The lower
(thin) curve in Figure 3
exhibits more frequent large oscillations,
due to more efficient energy transfer between the singly excited states.
The upper
(thick) curve in Figure 3 shows an increase,
relative to the corresponding
case of Figure 1,
in the magnitude of
oscillation, consistent with the fact that there is now more spectral
overlap of the excitation pulse with states having significant
amplitude on the higher-energy as well as the lower-energy chromophore.

In Figure 4, time-resolved anisotropies are shown for a case of two
identical chromophores, with the same excitation and detection conditions as
in the previous calculations.  The anisotropies can be seen to
reflect the fact that this is the case with the most efficient
excitation transfer.  The frequency of the large amplitude
oscillations is the highest
compared to Figures 1 and 3 in
the case of excitation 250 cm$^{-1}$ above the zero-zero electronic
transition frequency of the chromophores (thin curve).
Excitation at the zero-zero
electronic transition frequency of the chromophores (thick curve) shows
the largest extent of excitation transfer (even leading to a negative
anisotropy at t $\approx$ 330 fs)
compared to its counterparts
in Figures 1 and 3.

Figure 5 examines the dependence of the initial ($t=0$) anisotropy on
the strength of the excitation transfer coupling while the rest of the
parameters are kept constant. As the excitation
transfer coupling is increased, the initial anisotropy decreases because
the initial rapid decrease of the anisotropy, calculated before time
averaging according to Eq. (\ref{eq:e2.16}) (not shown), becomes more
rapid with increasing coupling, and keeping the time-resolution of the
detector constant results in observing a lower average anisotropy during the
$t=0$ detection window.
It is evident that for the case where the
timescale of exciton dynamics is shorter than the timescale of the
excitation and detection process, the high initial anisotropy will not be
observed.

The time-resolved anisotropy reflects the exciton dynamics of the
chromophore pair. It would be interesting to determine the extent to
which the
time-resolved anisotropy manifests the net transfer of excited state
population between the chromophores.\cite{rachsilbey}
The time-dependent site population difference is given by
$\langle \Psi(t)|P_1-P_2|\Psi(t)\rangle$, where $|\Psi(t)\rangle$ is the
state of the chromophore pair at time $t$, and $P_1$ an $P_2$ are given by
\begin{equation}
P_1 = \sum_{n_e,m_g} |eg \rangle \langle eg| \;
|n_e m_g \rangle \langle n_e m_g|  \label{eq:e4.p1}
\end{equation}
and
\begin{equation}
P_2 = \sum_{n_g,m_e} |ge \rangle \langle ge| \;
|n_g m_e \rangle \langle n_g m_e| \; . \label{eq:e4.p2}
\end{equation}
At times after the excitation pulse has subsided, the time-dependent
orientationally averaged population difference can be expressed by
\begin{eqnarray}
\langle \Psi(t)|P_1-P_2|\Psi(t)\rangle = \sum_{\gamma,\gamma^\prime}
\mu_{\alpha \gamma} \mu_{\alpha \gamma^\prime}
\overline{({\bf e}_{\alpha \gamma} \cdot {\bf e}_L)
({\bf e}_{\alpha \gamma^\prime} \cdot {\bf e}_L)} \nonumber \\
\times F(\varepsilon_{\gamma \alpha}-\Omega)
F(\varepsilon_{\gamma^\prime \alpha}-\Omega)
e^{iE_{\gamma,\gamma^\prime}t}
\langle\gamma|(P_1 - P_2)|\gamma^\prime\rangle
\; , \label{eq:e4.p1minp2}
\end{eqnarray}
where
\begin{equation}
F(\omega) = \sqrt{2 \pi} \tau_L \exp{(-\tau_L^2 \omega^2/2)} \; .
\label{eq:e4.Fofomega}
\end{equation}

Figure 6 shows the calculation of the expectation value of
$P_1(t)-P_2(t)$, normalized by the total excited state population,
$P_1(t)+P_2(t)$, for
the same parameters as in Figure 1.  Again, the initial state
of the chromophore
pair is taken to be $|\alpha\rangle = |gg\rangle|0_g0_g\rangle$.
The thick curve
shown in Figure 6a is the time-dependent site population difference
in the case of excitation centered at the zero-zero electronic transition
frequency of chromophore 1, the lower-energy chromophore, and the thin
curve is the corresponding time-dependent anisotropy.  In Figure 6b, the
thick curve is the
time-dependent site population difference for the excitation frequency
centered 100 cm$^{-1}$ above the zero-zero electronic transition frequency of
(the higher-energy)
chromophore 2, and the thin curve is the
again the corresponding time-resolved anisotropy.
The excitation condition
that leads to the site population difference curve in Figure 6a
places most of the
initial excitation on chromophore 1, as evidenced by the positive
initial value of
$\frac{\textstyle P_1(t)-P_2(t)}{\textstyle P_1(t)+P_2(t)}$.
The relatively small oscillations of
the upper (thick) curve indicate that most of the excitation remains on
the lower-energy chromophore. This is consistent with the anisotropy
results shown in the thin curve of Figure 6a, where the
time-resolved anisotropy obtained with the same excitation conditions
shows relatively small oscillations and high anisotropy.
Excitation centered 100 cm$^{-1}$ above the zero-zero electronic transition
frequency of the higher-energy chromophore,
resulting in a site population difference shown in the thick curve of
Figure 6b, also exhibits behavior consistent with
the excitation conditions. $\frac{\textstyle P_1(t)-P_2(t)}
{\textstyle P_1(t)+P_2(t)}$
is initially negative, indicating
that the excitation conditions place most
of the excitation on the higher-energy
chromophore; even though there is good
spectral overlap between the excitation pulse and states on both
chromophores,  the Franck-Condon factors are more favorable for the
excitation of the higher-energy chromophore for the coordinate
displacement between
ground and excited states that we have chosen.
The large oscillations about a relatively low value of the site
population difference indicate efficient net population transfer between the
two chromophores.  This behavior is also reflected in the large
oscillations seen in the corresponding time-dependent anisotropy
shown in the thin curve in Figure 6b.
In fact, the first several extrema in the population difference of
Figure
6b coincide with corresponding extrema in the anisotropy.  In both Figs.
6a and 6b, however, the anisotropies show high frequency structure
attributable to coherent vibrational motion, that is not as clearly
evident in the excitation probability difference.

\section*{5. Effects of Vibrational Relaxation and Dephasing}

In the previous sections we developed a model for an isolated
chromophore pair and
a detector in order to describe the time-resolved fluorescence
anisotropy following excitation with an ultrashort laser pulse.
The time-resolved anisotropies shown in Section 4 exhibited a similarity
in excitation wavelength dependence to the experimental measurements
made on APC and C-PC. \cite{xiedumets}  The experimental anisotropy
measurements did not, however, show evidence of the coherent
oscillations in the time-resolved anisotropy that were seen in the
calculations on isolated chromophore pairs.  The
surrounding medium undoubtedly plays a major role in destroying the
coherences between exciton states that give rise to the
oscillations seen in the calculations of Section 4.  In this
section, we include two aspects of the surrounding medium, namely,
vibronic energy relaxation and dephasing, and examine their effects
on the time-resolved anisotropy.

Our inclusion of vibrational relaxation and dephasing builds upon the
approach of Jean, Friesner and Fleming. \cite{jff}  In their work, Jean
{\em et al.} developed a quantum mechanical theory of photo-induced
electron transfer (see also the recent work of Pollard and Friesner on
the application of Redfield theory to multilevel systems\cite{pollard}
and work by Walsh and Coalson \cite{walsh}).
Using Redfield relaxation theory, they
incorporated vibrational relaxation through a coupling of the
quantum-mechanical reaction coordinate of an electron transfer system
to a thermal bath.  We
adopt a similar approach in our work, introducing a coupling of the
intra-chromophore vibrations to a thermal bath.

In Section 5A, we derive the form of the
time- and frequency-resolved signal in terms of the elements of the reduced
chromophore pair density matrix.
Then we introduce and describe the system-bath
coupling operator in Section 5B.  Section 5C explains some important
details of our Redfield calculations, and results of
numerical calculations of the time-resolved anisotropy including
vibrational relaxation and dephasing are presented in Section 5D.

{\bf 5A. Density Matrix Increment}

In deriving a density matrix expression for
the time- and frequency-resolved signal, the
notation of Section 4 will be used.  In order to calculate the excited
state population of the detector we need the contribution to the
density matrix that is second order in the electric field of the
excitation laser pulse and second order in the interaction with the
detector.  This contribution is given in the interaction picture by
\begin{eqnarray}
\tilde{\rho}^{(4)}(t) = \int_{t_0}^t d\tau \int_{t_0}^\tau d\tau_1
\int_{t_0}^{\tau_1} d\tau_2 \int_{t_0}^{\tau_2} d\tau_3  \:
\{ \widetilde{V}_d(\tau)\widetilde{V}_e(\tau_1)\rho_0
\widetilde{V}_e(\tau_3)\widetilde{V}_d(\tau_2)
\nonumber \\
\, + \, \widetilde{V}_d(\tau)\widetilde{V}_e(\tau_3)\rho_0
\widetilde{V}_e(\tau_2)\widetilde{V}_d(\tau_1)
\, + \, \widetilde{V}_d(\tau)\widetilde{V}_e(\tau_2)\rho_0
\widetilde{V}_e(\tau_3)\widetilde{V}_d(\tau_1)
\, + \, h.c.\} \: , \label{eq:e5.1}
\end{eqnarray}
where $\widetilde{V}_d(t)$ and $\widetilde{V}_e(t)$ are the previously defined
interaction Hamiltonians in the interaction picture, given by Eqs.
(\ref{eq:e2.12a1}) and (\ref{eq:e2.12a}),
and $\rho_0$ is the initial density matrix of the
entire system, which is assumed initially to take the factorizable form
\begin{equation}
\rho_0 = \rho_{cp} \; \rho_b \; |a\rangle \langle a| .
\label{eq:e5.2}
\end {equation}
$\rho_{cp} = |\alpha\rangle\langle\alpha|$ is the initial reduced
density matrix of the chromophore pair and
$\rho_b$ is the initial density matrix of the thermal bath.
$H_0$ now includes the Hamiltonian of the thermal bath, $H_{b}$, and the
chromophore pair-bath coupling, $H_{bcp}$, and is given by
\begin{equation}
H_0 = H_{cp} + H_d + H_b + H_{bcp} \; . \label{eq:e5.hsubzero}
\end{equation}
The precise form of $H_{bcp}$
will be developed in Section 5B.

If the time $t$ in Eq. (\ref{eq:e5.1}) is large enough so that excitation
and detection do not overlap, and the density matrix increment is
defined as
$\Delta \widetilde{\rho}^{(4)}(t)$ as
\begin{equation}
\Delta \widetilde{\rho}^{(4)}(t) \equiv
\widetilde{\rho}^{(4)}(t+\frac{\Delta t}{2}) -
\widetilde{\rho}^{(4)}(t-\frac{\Delta t}{2}) \; , \label{eq:e5.inc}
\end{equation}
then the detected signal can be
calculated  with the help of
\begin{equation}
\Delta \widetilde{\rho}^{(4)}(t)
\: =  \int_{t-\frac{\Delta t}{2}}^{t+\frac{\Delta t}{2}}
d\tau \int_{t_0}^\tau d\tau_1
\int_{t_0}^{\infty} d\tau_2 \int_{t_0}^{\infty} d\tau_3  \:
\widetilde{V}_d(\tau)\widetilde{V}_e(\tau_3)\rho_0
\widetilde{V}_e(\tau_2)\widetilde{V}_d(\tau_1) + h.c.  \label{eq:e5.3}
\end{equation}
We arrive at Eq. (\ref{eq:e5.3}) by noticing that when excitation and
detection do not overlap, the first term on the right-hand side of Eq.
(\ref{eq:e5.1}) does not contribute to Eq. (\ref{eq:e5.3}), and the
upper limit of integration in the integral over $\tau_2$
can be replaced by infinity.

The integrand in Eq. (\ref{eq:e5.3}) can be simplified as follows.
First, we make an assumption that the effects of the bath are
negligible on the timescale of the duration of the laser pulse. That
approximation allows us to neglect the
system-bath coupling Hamiltonian in the terms
$\widetilde{V}_e(\tau_3)$ and $\widetilde{V}_e(\tau_2)$,
and Eq. (\ref{eq:e5.3}) becomes
\begin{eqnarray}
\Delta \widetilde{\rho}^{(4)}(t) \;
\cong \; \int_{t-\frac{\Delta t}{2}}^{t+\frac{\Delta t}{2}}
d\tau \int_{t_0}^\tau d\tau_1
\frac{E_0^2}{4}\sum_{\gamma,\gamma^\prime} \mu_{\alpha \gamma}
\mu_{\alpha \gamma^\prime}
({\bf e}_{\alpha \gamma} \cdot {\bf e}_L)
({\bf e}_{\alpha \gamma^\prime} \cdot {\bf e}_L)  \nonumber \\
\times F(\varepsilon_{\gamma \alpha}-\Omega)
F(\varepsilon_{\gamma^\prime \alpha}-\Omega)
\widetilde{V}_d(\tau) \; |\gamma \rangle \langle \gamma^{\prime}| \;
\rho_b \; |a\rangle \langle a|
\widetilde{V}_d(\tau_1) \: + h.c. \label{eq:e5.4}
\end{eqnarray}
where
$F(\varepsilon_{\gamma \alpha}-\Omega)$ has been previously defined by
Eq. (\ref{eq:e4.Fofomega}).
We next write out explicitly the
operators $\widetilde{V}_d(\tau)$ and $\widetilde{V}_d(\tau_1)$ in Eq.
(\ref{eq:e5.4}) and take the expectation value of
$\Delta \widetilde{\rho}^{(4)}(t)$ in state $|b\rangle
|\alpha^{\prime}\rangle$.  Since the Hamiltonian of the detector
commutes with the rest of the operators in $H_0$, we replace $H_d$ with
the corresponding eigenenergy.  Integrating the resulting expression
over the frequency profile of the detector according to
Eq. (\ref{eq:e2.15}) yields
\begin{eqnarray}
\int_{-\infty}^{\infty} d\varepsilon_d
D(\varepsilon_d-\overline{\varepsilon}_d)
\langle b|\langle\alpha^{\prime}|\Delta \tilde{\rho}^{(4)}(t)
|\alpha^{\prime}\rangle|b\rangle \; = \; \nonumber \\
\frac{E_0^2 \eta^2 \mu_d^2}{4}
\sum_{\gamma,\gamma^{\prime},\beta,\beta^{\prime}}
\mu_{\alpha \gamma} \mu_{\alpha \gamma^{\prime}}
\mu_{\alpha^{\prime} \beta} \mu_{\alpha^{\prime} \beta^{\prime}}
({\bf e}_{\alpha \gamma} \cdot {\bf e}_L)
({\bf e}_{\alpha \gamma^{\prime}} \cdot {\bf e}_L)
({\bf e}_{\alpha^{\prime} \beta} \cdot {\bf e}_d)
({\bf e}_{\alpha^{\prime} \beta^{\prime}} \cdot {\bf e}_d) \nonumber \\
\times\int_{t-\frac{\Delta t}{2}}^{t+\frac{\Delta t}{2}}
d\tau \int_{t_0}^\tau d\tau_1
F(\varepsilon_{\gamma \alpha}-\Omega)
F(\varepsilon_{\gamma^{\prime} \alpha}-\Omega)
e^{i\overline{\varepsilon}_d (\tau - \tau_1)}
e^{-(\tau -\tau_1)^2 \Delta^2/4} \nonumber \\
\times\langle \alpha^{\prime}|e^{iH_0^{\prime} \tau}|\alpha^{\prime} \rangle
\langle \beta | e^{-iH_0^{\prime}\tau}|\gamma \rangle \langle
\gamma^{\prime}| \; \rho_b \;
e^{iH_0^{\prime} \tau_1} | \beta^{\prime}\rangle \langle
\alpha^{\prime} | e^{-iH_0^{\prime}\tau_1}|\alpha^{\prime} \rangle
\: + h.c. \: , \label{eq:e5.6}
\end{eqnarray}
where $D(\varepsilon_d-\overline{\varepsilon}_d)$
denotes the frequency window of
the detector, given by Eq. (\ref{eq:e2.15D}), and
the prime on $H_0^{\prime}$ indicates that $H_d$ has been removed.
At this stage we make a similar assumption regarding the
coupling of the chromophore pair to the bath during detection
as we made during the excitation pulse, namely, that the bath has
a negligible effect on the chromophore pair on the timescale of the
inverse of the frequency width of the detector, $\Delta^{-1}$.  That
assumption
allows us to make the following approximate replacement in Eq.
(\ref{eq:e5.6})
\begin{equation}
\langle \beta | e^{-iH_0^{\prime}\tau}|\gamma \rangle \langle
\gamma^{\prime}| \; \rho_b \;
e^{iH_0^{\prime} \tau_1} | \beta^{\prime}\rangle \cong
\langle \beta | e^{-iH_0^{\prime}\tau}|\gamma \rangle \langle
\gamma^{\prime}| \; \rho_b \;
e^{iH_0^{\prime} \tau} | \beta^{\prime}\rangle
e^{-i(H_b+E_{\beta^{\prime}})(\tau-\tau_1)} \: . \label{eq:e5.7}
\end{equation}
Using Eq. (\ref{eq:e5.7}) in Eq. (\ref{eq:e5.6}) and taking the trace
over the bath results in the final expression
for the population of the state $|\alpha^\prime \rangle |b\rangle$:
\begin{eqnarray}
Tr_b\{
\int_{-\infty}^{\infty} d\varepsilon_d
D(\varepsilon_d-\overline{\varepsilon}_d)
\langle b|\langle\alpha^{\prime}|\Delta \tilde{\rho}^{(4)}(t)
|\alpha^{\prime}\rangle|b\rangle \} \; =\; \nonumber \\
\frac{E_0^2 \eta^2}{4} \sum_{\gamma,\gamma^\prime,\beta,\beta^\prime}
\mu_{\alpha \gamma} \mu_{\alpha \gamma^\prime}
\mu_{\alpha^\prime \beta} \mu_{\alpha^\prime \beta^\prime}
({\bf e}_{\alpha \gamma} \cdot {\bf e}_L)
({\bf e}_{\alpha \gamma^\prime} \cdot {\bf e}_L)
({\bf e}_{\alpha^{\prime} \beta} \cdot {\bf e}_d)
({\bf e}_{\alpha^{\prime} \beta^\prime} \cdot {\bf e}_d) \nonumber \\
\times\int_{t-\frac{\Delta t}{2}}^{t+\frac{\Delta t}{2}}
d\tau \int_{t_0}^\tau d\tau_1
F(\varepsilon_{\gamma \alpha}-\Omega)
F(\varepsilon_{\gamma^{\prime} \alpha}-\Omega)
e^{-i(\varepsilon_{\beta^\prime \alpha^\prime} -
\overline{\varepsilon}_d)(\tau - \tau_1)}
e^{-(\tau -\tau_1)^2 \Delta^2/4} \nonumber \\
\times Tr_b\{ \langle \beta | e^{-iH_0^{\prime}\tau}|\gamma \rangle \langle
\gamma^{\prime}| \; \rho_b \;
e^{iH_0^{\prime} \tau} | \beta^{\prime}\rangle \} + c.c. \label{eq:e5.8}
\end{eqnarray}
Making approximations that neglect the effects of the bath during
the excitation pulse and the detection time window confines the
effects of the bath to
$Tr_b\{ \langle \beta | e^{-iH_0^\prime\tau}|\gamma \rangle \langle
\gamma^{\prime}| \; \rho_b \;
e^{iH_0^\prime \tau} | \beta^\prime\rangle \}$. In order to evaluate the
latter quantity we must now develop  a model for the bath and the form
of the bath-chromophore pair interaction.

{\bf 5B. Bath-Chromophore Pair Interaction}

Before proceeding farther, we make the following definition regarding the
system and the bath.  The degrees of freedom that we choose to treat
explicitly are the two intra-chromophore modes, referred to as $Q_1$ and
$Q_2$.  In general, we assume that any optically active collective
coordinates (i.e. those whose equilibrium values are displaced in the
excited electronic states) would be treated on the same explicit footing as
$Q_1$ and $Q_2$.
Thus the bath modes are defined to be those collective
coordinates that are not displaced when an optical transition takes
place.

Consistently with our assumption of the factorizable form, Eq.
(\ref{eq:e5.2}) of the total initial density matrix in
the electronic ground state, we neglect bath-chromophore pair
interaction entirely prior to electronic excitation.
We expand the system-bath coupling operator $V_{eg}(Q_1,Q_2,q)$
for the singly excited state $|eg\rangle$
about the
equilibrium positions of $Q_1$ and $Q_2$ in that state, denoted by
$\Delta^{eg}_1$ and $\Delta^{eg}_2$, respectively, and let $q$ denote the
bath coordinates.
Expanding
$V_{eg}(Q_1,Q_2,q)$ to first order in the displacement of $Q_1$ and $Q_2$
yields
\begin{eqnarray}
V_{eg}(Q_1,Q_2,q) \cong V_{eg}(\Delta_1^{eg},\Delta_2^{eg},q) +
(Q_1-\Delta_1^{eg}) \frac {\partial
V_{eg}(\Delta_1^{eg},\Delta_2^{eg},q)}{\partial Q_1} + \nonumber \\
(Q_2-\Delta_2^{eg}) \frac {\partial
V_{eg}(\Delta_1^{eg},\Delta_2^{eg},q)}{\partial Q_2} .  \label{eq:e5.9}
\end{eqnarray}
The $q$-dependence of the first term
on the right-hand side of Eq. (\ref{eq:e5.9}) would, in general,
contribute to optical dephasing. Considering $q$ to be a single degree of
freedom for purposes of illustration, we expand the first term on the
right-hand-side of Eq. (\ref{eq:e5.9}) in powers of $q$ about its
equilibrium position, which is chosen to be zero (and by
definition, to be the same in the ground and both excited electronic
states):
\begin{eqnarray}
V_{eg}(\Delta_1^{eg},\Delta_2^{eg},q) \cong V_{eg}(\Delta_1^{eg},
\Delta_2^{eg},0) +
\frac{q^2}{2}
\frac{\partial^2 V_{eg}(\Delta_1^{eg},\Delta_2^{eg},0)}{\partial q^2} + ...
\label{eq:e5.10}
\end{eqnarray}
where the term linear in $q$ vanishes.  Terms quadratic,
cubic, etc. in $q$ represent changes in frequency, anharmonicity, etc. of
that bath mode following optical excitation of chromophore 1. Because
a collective coordinate of the type $q$ will generally be of low
frequency (long wavelength), its change in frequency, anharmonicity,
etc. will be assumed to be negligibly small.  The leading term in Eq.
(\ref{eq:e5.9}) is thereby taken to be sensibly $q$-independent as well
as independent of the intramolecular coordinates; it can be absorbed
into the nuclear Hamiltonian $H_{eg}$.
Hence, we arrive at a
working model in which all optically active collective coordinates are
treated explicitly, the remaining (optically inactive) coordinates are
relegated to the bath, and a phenomenological treatment of electronic
dephasing need not be introduced.  The remaining $q$-dependence in Eq.
(\ref{eq:e5.9}) resides in terms linear in $(Q_1-\Delta_1^{eg})$ and
$(Q_2-\Delta_2^{eg})$.  These $q$-dependent terms govern bath-induced
vibrational relaxation in the site state $|eg\rangle$.

In order to avoid unnecessary complications, we make an additional
assumption that
$Q_1$ and $Q_2$ couple to orthogonal combinations of bath
coordinates, denoted by $q_1$ and $q_2$, respectively.  The system-bath
coupling becomes
\begin{eqnarray}
V_{eg}(Q_1,Q_2,q_1,q_2) & = & V_{eg}^{(1)}(Q_1,q_1)+V_{eg}^{(2)}(Q_2,q_2)
 \nonumber \\
& \cong &
(Q_1-\Delta^{eg}_1) \frac {\partial
V_{eg}^{(1)}(\Delta_1^{eg},q_1)}{\partial Q_1}
 +  (Q_2-\Delta^{eg}_2) \frac {\partial
V_{eg}^{(2)}(\Delta_2^{eg},q_2)}{\partial Q_2} .  \label{eq:e5.11}
\end{eqnarray}
There is an analogous expression for $V_{ge}(Q_1,Q_2,q_1,q_2)$.
Since we are neglecting frequency shifts of the bath modes, it is
consistent to make linear approximations to the derivatives in Eq.
(\ref{eq:e5.11}):
\begin{equation}
\frac{\partial V_{eg}^{(1)}(\Delta_1^{eg},q_1)}
{\partial Q_1} \cong q_1
\frac{\partial^2 V_{eg}^{(1)}(\Delta_1^{eg},0)}
{\partial q_1 \partial Q_1} \equiv f_lq_1
\label{eq:e5.feq1}
\end{equation}
and similarly for
$\frac{\textstyle \partial V_{eg}^{(2)}(\Delta_2^{eg},q_2)}
{\textstyle \partial Q_2}$.
In the end, assuming the bilinear coupling constants
to be the same for both pairs of intramolecular and bath modes in both
electronic excited states allows us to write simply
\begin{eqnarray}
H_{bcp} & = & |eg\rangle \langle eg| \{(Q_1-\Delta_1^{eg})f_lq_1 +
(Q_2-\Delta_2^{eg})f_lq_2)\} \nonumber \\
& + & |ge\rangle \langle ge| \{(Q_1-\Delta_1^{ge})f_lq_1 +
(Q_2-\Delta_2^{ge})f_lq_2)\} \; . \label{eq:e5.hbcp}
\end{eqnarray}
In practice, we hold it that electronically exciting one chromophore
leaves an intramolecular mode on the other undisplaced, so that
$\Delta^{eg}_2 = \Delta^{ge}_1 = 0$.

{\bf 5C. Vibronic Relaxation}

In Eq. (\ref{eq:e5.8}) for the detector-frequency-weighted populations
in the states $|\alpha^\prime \rangle |b\rangle$,
the effects of the bath
have been confined to the terms
$Tr_b\{ \langle \beta | e^{-iH_0^{\prime}\tau}|\gamma \rangle \langle
\gamma^{\prime}| \; \rho_b \;
e^{iH_0^{\prime} \tau} | \beta^{\prime}\rangle \}$, which is the
$\beta \beta^{\prime \, th}$ element of the reduced chromophore pair
operator $\sigma (t)$ obtained by propagating
$\sigma(0) \equiv |\gamma \rangle \langle \gamma^\prime | \rho_b$
under $H_0^\prime$ and tracing over the bath.  It is convenient to
propagate the elements of the initial density matrix of the system term
by term in order to facilitate the orientational averaging.
Since the system-bath coupling is assumed to be weak, we can follow the recent
work of Jean and co-workers and use Redfield theory
to calculate the time development of the operators $\sigma (t)$.

Since Redfield theory is most naturally implemented here
in the exciton (i.e.
eigenstate) representation of the chromophore pair Hamiltonian,
the system-bath coupling must be transformed from the site
representation, in which the matrix elements of $H_{bcp}$,
Eq. (\ref{eq:e5.hbcp}), between site
vibrational quantum number states are readily
calculated, to the
exciton representation.  The system-bath coupling in the exciton
representation connects various exciton states and ultimately leads to
population and phase relaxation among the exciton states.

The Redfield equations of motion for the reduced operators $\sigma(t)$
have the form \cite{jff,pollard,wertheimer}
\begin{equation}
\dot{\sigma}_{\beta \beta^{\prime}}(t) =
-i\varepsilon_{\beta \beta^{\prime}}\sigma_{\beta \beta^{\prime}}(t)
+ \sum_{\kappa, \kappa^{\prime}}
R_{\beta \beta^{\prime},\kappa \kappa^{\prime}} \:
\sigma_{\kappa \kappa^{\prime}}(t) \label{eq:e5.redfield}
\end{equation}
and the Redfield tensor $R_{\beta
\beta^\prime,\kappa \kappa^\prime}$ is specified by
\begin{eqnarray}
R_{\beta \beta^{\prime},\kappa \kappa^{\prime}} =
\int_0^\infty dt \{
\langle [H_{bcp}(t)]_{\kappa^\prime \beta^\prime}
[H_{bcp}(0)]_{\beta \kappa}\rangle
 e^{i\varepsilon_{\kappa \beta}t}
+ \langle [H_{bcp}(0)]_{\kappa^\prime \beta^\prime}
 [H_{bcp}(t)]_{\beta \kappa}(t)\rangle
 e^{i\varepsilon_{\beta^\prime \kappa^\prime}t} \nonumber \\
-\delta_{\kappa^\prime \beta^\prime}
\sum_\xi \langle [H_{bcp}(t)]_{\beta \xi}
[H_{bcp}(0)]_{\xi \kappa}\rangle
e^{i\varepsilon_{\kappa \xi}t}
- \delta_{\kappa \beta}
\sum_\xi \langle [H_{bcp}(0)]_{\kappa^\prime \xi}
[H_{bcp}(t)]_{\xi \beta^\prime}\rangle
e^{i\varepsilon_{\xi \kappa^\prime}t} \} \label{eq:e5.rtensor}
\end{eqnarray}
where $[H_{bcp}(t)]_{\beta \kappa} \equiv e^{iH_bt}
[H_{bcp}]_{\beta \kappa}e^{-iH_bt}$
and $\langle\cdots\rangle \equiv Tr_b\{\cdots \rho_b\}$.

The bath-chromophore pair interaction in
Eq. (\ref{eq:e5.hbcp}) is a sum of
products of chromophore pair and bath operators, and the transformation of
the system-bath coupling operator to the exciton representation leaves
the bath operators untouched.  Therefore, bath-chromophore pair interactions in
the exciton representation can readily be expressed as a sum of products of
chromophore pair and bath operators, and we adopt Pollard and Friesner's
notation for the Redfield tensor which is appropriate to such
a form \cite{pollard}
\begin{eqnarray}
R_{\beta \beta^{\prime},\kappa \kappa^{\prime}} =
\sum_{a=1}^{2} [(G_a)_{\kappa^\prime \beta^\prime} (G_a)_{\beta \kappa}
\{(\Theta_{aa}^+)_{\beta \kappa} + (\Theta_{aa}^-)_{\kappa^\prime
\beta^\prime}\} \nonumber \\
- \delta_{\kappa^\prime \beta^\prime} \sum_\xi
(G_a)_{\beta \xi} (G_a)_{\xi \kappa} (\Theta_{aa}^+)_{\xi \kappa}
- \delta_{\beta \kappa} \sum_\xi
(G_a)_{\kappa^\prime \xi} (G_a)_{\xi \beta^\prime}
(\Theta_{aa}^-)_{\kappa^\prime \xi}] \: , \label{eq:e5.pollardtens}
\end{eqnarray}
where (see Eq. (\ref{eq:e5.hbcp}))
\begin{eqnarray}
G_1 = |eg \rangle \langle eg|(Q_1-\Delta_1^{eg})f_l +
|ge \rangle \langle ge|Q_1f_l \nonumber \\
G_2 = |eg \rangle \langle eg|Q_2 f_l +
|ge \rangle \langle ge|(Q_2-\Delta_2^{ge})f_l \; , \label{eq:e5.G1G2}
\end{eqnarray}
and
$(\Theta_{aa}^\pm)_{\beta \delta}$ are given by
\begin{equation}
(\Theta_{aa}^\pm)_{\beta \delta} = \int_0^\infty d\tau
e^{-i\varepsilon_{\beta \delta}\tau} \langle q_a(\pm \tau)q_a\rangle
\: . \label{eq:e5.theta1}
\end{equation}
The expression for the Redfield tensor given by Pollard and Friesner
\cite{pollard} also includes contributions from cross-correlations of different
sets of bath coordinates, but since we assume that $q_1$ and $q_2$ are
completely uncorrelated, the terms involving
cross-correlations between $q_1$ and $q_2$ vanish.

Neglecting the imaginary part of
$(\Theta_{aa}^\pm)_{\beta \delta}$, Eq. (\ref{eq:e5.theta1}) can be
rewritten as
\begin{eqnarray}
(\Theta_{aa}^\pm)_{\beta \delta} & \cong &
\frac{1}{2} \int_{-\infty}^\infty
d\tau e^{-i\varepsilon_{\beta \delta}\tau} \langle q_a(\pm \tau)q_a\rangle
\label{eq:e5.theta2} \\
& \cong & e^{\mp \beta \varepsilon_{\beta \delta}\tau}
\frac{1}{2} \int_{-\infty}^\infty
d\tau e^{-i\varepsilon_{\beta \delta}\tau} \langle q_a q_a(\pm \tau)\rangle
\: . \label{eq:e5.theta3}
\end{eqnarray}
Equations (\ref{eq:e5.theta2}) and (\ref{eq:e5.theta3}) can be combined to
obtain $(\Theta_{aa}^\pm)_{\beta \delta}$ in terms of the symmetrized
correlation function,
\begin{equation}
(\Theta_{aa}^\pm)_{\beta \delta} =
\frac{1}{1+e^{\pm \beta \varepsilon_{\beta \delta}}}
\int_{-\infty}^\infty
d\tau e^{-i\varepsilon_{\beta \delta}\tau} \frac{1}{2}
\{\langle q_a q_a(\pm \tau)\rangle
+ \langle q_a(\pm \tau) q_a\rangle\} \; , \label{eq:e5.theta3a}
\end{equation}
Pollard and Friesner have emphasized that whatever its shortcomings, the
neglect of the imaginary part of $(\Theta_{aa}^\pm)_{\beta \delta}$ does
not undermine detailed balance. In fact,
the symmetrized correlation function in Eq. (\ref{eq:e5.theta3a}) can
be replaced by any approximate function that is real and
even with respect to time.  We introduce a mean-squared fluctuation and bath
correlation time by hypothesizing that
\begin{equation}
\frac{1}{2}(\langle q_a q_a(\pm \tau)\rangle +
\langle q_a(\pm \tau) q_a\rangle =
\langle q_a^2 \rangle e^{-|\tau|/\tau_c}
\: , \label{eq:e5.classcor}
\end{equation}
Using Eq. (\ref{eq:e5.classcor}) in  Eq. (\ref{eq:e5.theta3a}) leads to
\begin{eqnarray}
(\Theta_{aa}^\pm)_{\beta \delta} & \cong &
\frac{1}{1+e^{\pm \beta \varepsilon_{\beta \delta}}}
\int_{-\infty}^\infty d\tau
e^{-i\varepsilon_{\beta \delta}\tau} \langle q_a^2 \rangle e^{-|\tau|/\tau_c}
\nonumber \\
& \cong & \frac{1}{1+e^{\pm \beta \varepsilon_{\beta \delta}}} \cdot
\frac{2/\tau_c}{{1/\tau_c^2 + \varepsilon_{\beta \delta}^2}}
\langle q_a^2 \rangle \: . \label{eq:5.theta4}
\end{eqnarray}
When $\tau_c$ is short compared to all relevant bohr periods in the
system, we obtain
\begin{equation}
(\Theta_{aa}^\pm)_{\beta \delta} =
\frac{1}{1+e^{\pm \beta \varepsilon_{\beta \delta}}}
(2\tau_c \langle q_a^2 \rangle) \: . \label{eq:e5.theta5}
\end{equation}

This panoply of assumptions and approximations allows us to specify the
entire Redfield tensor by choosing a single parameter. \cite{pollard}
In the absence of excitation transfer, the Redfield tensor element
governing the rate of population decay from $|eg\rangle|1_e0_g\rangle$
($\equiv |1\rangle$)
to $|eg\rangle|0_e0_g\rangle$ ($\equiv |0\rangle$) is
\begin{equation}
R_{00,11} = (G_1)_{10}(G_1)_{01}\{\frac{2}{1+e^{-\beta\omega}} \cdot
2\tau_c \langle q_1^2\rangle \} \; , \label{eq:e5r0011}
\end{equation}
where
$(G_1)_{10}=(G_1)_{01}=f_l/\sqrt{2\omega}$,
and we can therefore choose the product of $f_l^2$ and
$\tau_c \langle q_1^2\rangle$ that yields a specified vibrational
population decay rate from $|eg\rangle|1_e0_g\rangle$ to
$|eg\rangle|0_e0_g\rangle$.

In our calculations we make the secular approximation, namely, we keep
only those elements of the Redfield tensor
$R_{\beta \beta^{\prime},\kappa \kappa^{\prime}}$
in which the element
is not significantly smaller than the frequency mismatch
$|\omega_{\beta\beta^\prime}-\omega_{\kappa\kappa^\prime}|$.

{\bf 5D. Anisotropy Decays}

Figure 7 shows calculated anisotropies for a chromophore pair with
molecular parameters,
laser pulse duration, and detector time and frequency resolution all
the same as those used for the calculations of Figure 1.
The initial state of the chromophore pair is
$|gg\rangle|0_g0_g\rangle$ and the bath is at 300 K.
The population relaxation rate (in the absence of energy transfer)
from $|eg\rangle|1_e0_g\rangle$ to
$|eg\rangle|0_e0_g\rangle$
is chosen to be 1.5 ps$^{-1}$ and for times less than
150 fs (overlapping laser pulse and detection window)
the dissipative effects of the bath are not included.
This is done in order to properly include the contributions
to the fluorescence rate present when the excitation pulse and the
detection window overlap.
We combine the instantaneous fluorescence rate obtained from an
isolated chromophore pair for times less than 150 fs with the rate
including dissipation for times longer than 150 fs.  The combined rate
is then integrated and used to calculate the time-resolved signal.
Since the combined rate is composed of two different calculations, there
is a discontinuity in the combined rate, leading to a slight
discontinuity in the slope of the short-time anisotropy.

The dissipative effects of the bath become most evident after
about 0.75 ps, where the calculated anisotropies in Figures
1 and 7 cease to resemble each other.
After 1.5 ps, the absence of
oscillations in the time-resolved anisotropies of Figure 7 suggests that
the coherences between
the exciton states are almost completely destroyed by the bath, and the
slow decay of the anisotropies beyond 1.5 ps results from the populations
in the reduced density matrix of the chromophore pair approaching their
equilibrium values at 300 K.  Our claim that the slow monotonic decay of
the time-resolved anisotropies beyond 1.5 ps is due to the thermal
equilibration of populations is supported by the calculations of the
site excitation probability presented below.
The lower (thin) curve in Figure 7 shows a more significant decay in the
1.5-4 ps region than the upper (thick) curve because the excitation
conditions that result in the anisotropy shown by the lower
(thin) curve prepare a higher-lying superposition of states that
subsequently decays into states localized on
(the lower-energy) chromophore 1.
Excitation conditions resulting in the anisotropy shown by the upper
(thick) curve prepare a superposition of states fairly well
localized on
the lower-energy chromophore, and therefore, population
relaxation does not result in a significant decay of the anisotropy in
the 2-4 ps region.

Figure 8 presents a calculation of the time-resolved
anisotropy designed to check whether the time-resolved anisotropy behaves in
a qualitatively similar manner when higher-lying thermally populated
initial ground states are included.
The contributions from the
lowest three thermally populated levels
$|gg\rangle|0_g0_g\rangle$,
$|gg\rangle|1_g0_g\rangle$, and
$|gg\rangle|0_g1_g\rangle$
have been included in Figure 8.  The contributions to the anisotropy from
$|gg\rangle|1_g0_g\rangle$ and
$|gg\rangle|0_g1_g\rangle$ are qualitatively similar to the
$|gg\rangle|0_g0_g\rangle$, and
calculated anisotropies in Figure 8 show a
similar behavior to those shown in Figure 7.
In the case of excitation 100 cm$^{-1}$ above the zero-zero
transition frequency of the higher-energy chromophore, there is a
perceptible smoothing of the coherent oscillations at early times,
compared to its counterpart in Figure 7.

The initial anisotropy in Fig. 8 still remains above 0.4 even with the
inclusion of higher-lying thermally populated initial ground states.
For example, consider the initial state $|gg\rangle|1_g0_g\rangle$.
The short-time excitation is dominated by the strongest Franck-Condon
transitions and occurs to the states $|ge\rangle|1_g0_e\rangle$ and
$|eg\rangle|1_e0_g\rangle$.
(Recall that for very short delays both the excitation pulse and
the detection window are effectively shortened, increasing their
effective bandwidths.)
Because each of these singly-excited states involves excitation on a
different chromophore,
the emission from these states back to $|gg\rangle|1_g0_g\rangle$
is the type that leads to a larger than 0.4 initial anisotropy.

Figure 9 shows a calculation of the time-resolved anisotropy,
including vibrational relaxation, for the
same case of a pair of identical chromophores and the same
excitation and detection
parameters used in Figure 4.  The vibrational population decay
rate of 1.5 ps$^{-1}$ from $|eg\rangle |1_e0_g\rangle$ to
$|eg\rangle |0_e0_g\rangle$
in the absence of energy transfer was again used in this calculation. The
overall qualitative behavior of the two anisotropy curves is
in some ways similar to
their counterparts in Figure 7.  However, the slow decay in the 2-4 ps
range seen in the anisotropy curves in Figure 7 is absent in the
corresponding curves in Figure 9.  This is consistent with the fact that
there can be no net population transfer
between two identical
chromophores.

The net site excitation probability difference $P_1(t)-P_2(t)$,
given by Eq. (\ref{eq:e4.p1minp2}) for the case of the isolated
chromophore pair, can also be calculated with the inclusion of
dissipative effects of the bath.  The net site population
difference can be written in terms of the reduced density matrix of
the chromophore pair
\begin{equation}
P_1(t)-P_2(t) = Tr_{cp}\{(P_1-P_2) Tr_b\{e^{-iH_0^\prime t}\rho_{cp}(0)\rho_b
e^{iH_0^\prime t}\}\} \; , \label{eq:e5.p1minp2diss}
\end{equation}
where
$H_0^\prime = H_{cp} + H_{b} + H_{bcp}$, and $Tr_{cp}\{\cdots\}$ denotes
the trace over the states of the chromophore pair.
The initial density matrix of the chromophore pair, neglecting the
effects of the bath during the excitation pulse, is given by
\begin{equation}
\rho_{cp}(0) = \sum_{\gamma, \gamma^\prime} |\gamma \rangle \langle
\gamma^\prime |
\mu_{\alpha \gamma} \mu_{\alpha \gamma^\prime}
\overline{({\bf e}_{\alpha \gamma} \cdot {\bf e}_L)
({\bf e}_{\alpha \gamma^\prime} \cdot {\bf e}_L)}
 F(\varepsilon_{\gamma \alpha}-\Omega)
F(\varepsilon_{\gamma^\prime \alpha}-\Omega) \label{eq:e5.rho0}
\end{equation}
Calculation of $P_1(t)-P_2(t)$ for the system parameters and excitation
conditions identical to those used for obtaining the anisotropies in
Figure 7 are shown in Figure 10.
The upper (thick) curve in Figure 10 is initially similar to
the thick curve in Figure 6a, where the effects of the bath were not
included.  $P_1(t)-P_2(t)$ for the isolated chromophore pair
is oscillatory but remains close to its initial value. When the bath is
included, the oscillations are damped out and the excitation probability
difference
tends to its equilibrium value.  The behavior of the excitation
probability difference shown by
the lower (thin) curve of Figure 10 is even more dramatic.
The large oscillations
present in its analog in Figure 6b are completely damped out by 2 ps,
indicating that coherent net population transfer has stopped.  Both
curves in Figure 10 approach the same steady-state value of 0.34,
consistent with thermal equilibration at 300 K.

\section*{6. Conclusions}
We have presented calculations of time-resolved fluorescence anisotropy
from a chromophore pair undergoing excitation transfer following
excitation with an ultrashort laser pulse.  An intramolecular mode for
each chromophore was explicitly included in our model,
in order to understand the effects of vibrational structure on the
anisotropy.
We have included
a model detector in order to take realistic account of
the limitations imposed by the
time and frequency resolution of the experimental detection process.

Choosing parameters for our model system to correspond
to the degree possible to known
parameters in closely-coupled chromophore pairs in C-PC, we have
investigated the excitation wavelength dependence of the
time-resolved anisotropy, observing behavior qualitatively similar to
the experimental findings. \cite{xiedumets}  Also, using a simplified
version of our model that can be treated analytically,
we have shown that the initial values of the
anisotropy are highly dependent on the ground state(s) into which the
chromophore is emitting.

The inclusion of vibrational structure is shown to play a
major role in determining the form of the time-resolved anisotropy.
Several effects are observed that are not present if each chromophore is
treated as an electronic 2-level system without
nuclear degrees of freedom. First, the
presence of multiple vibrational levels in the excited state of each
chromophore introduces excitation-wavelength dependence
that is significantly different from that found in a pair of coupled
two-level systems.  Second, in a chromophore pair
lacking vibrational structure,
coherent energy transfer requires a pulse bandwidth in excess of the
bare exciton splitting, which is greater than the site energy
difference.  On the other hand,
when each chromophore has a ladder of
excited vibronic states, preparation of a superposition of close-lying
vibronic states with amplitude on both chromophores can sometimes be
accomplished with a pulse bandwidth less than the difference between the
two sites in zero-zero transition frequency.
Finally, the vibrational levels in the ground electronic state selected
by the detection bandwidth
play a significant role in determining the effects of quantum mechanical
interference on the
time-resolved anisotropy.

We have reported Redfield theory calculations of the anisotropy, which
included coupling of the intramolecular coordinates to a surrounding
medium.
When the effects of a thermal bath were included,
the coherent oscillations in the anisotropy remained at short times but
were seen to
decay prior to the anisotropy reaching its steady-state value.
It is interesting to note, however, that evidence for the coherent
oscillations, which are present in our calculations including vibronic
relaxation, seems to be lacking in the experimental anisotropies
measured for APC and C-PC by Xie {\em et al.} \cite{xiedumets}
Failure to sum over all vibrational levels initially thermally populated
at 300 K causes the coherent oscillations in the anisotropy to be
exaggerated.
It can also be anticipated that addition of multiple
optically active intramolecular and/or collective modes would tend to
obscure the oscillatory contribution to the calculated signals.
Furthermore, the coherent oscillations in the anisotropy can be degraded
by the presence of inhomogeneous broadening in the zero-zero
frequency offset between the two chromophores. Even if the zero-zero transition
energies in the two chromophores are perfectly correlated, the inhomogeneous
broadening of the zero-zero transition energies will have the effect of
creating a distribution of different excitation and detection
conditions, which could also tend to mask the coherent anisotropy oscillations.

In very recent work, Jean \cite{jean} has reported
calculations on fluorescence
emission from an electron transfer system including vibronic relaxation
via Redfield theory and a treatment of the frequency-dependent
instantaneous emission rate essentially equivalent to our model
detection apparatus.  He reports the interesting result that under some
circumstances, the instantaneous emission rate can accurately reflect
the extent of electron transfer.

\section*{Acknowledgments}
We thank Sunney Xie, Mei Du, and Graham Fleming for many helpful
discussions and suggestions.  We also thank John Jean and Eric Hiller
for their advice regarding the implementation
of Redfield theory, and Lowell Ungar, Tim Smith, Stephen Bradforth
and Sandy Rosenthal for their
comments on the manuscript. This
work was supported by the National Science Foundation and the Camille
and Henry Dreyfus Teacher-Scholar Award Program.

\section*{Appendix: Orientational Averaging}
We need to obtain the average over the products of four direction
cosines in Eq.
(\ref{eq:e3.7}), where
${\bf e}_+$ and ${\bf e}_-$ are unit vectors in the molecule-fixed
frame, ${\bf e}_L$ and ${\bf e}_d$ are unit vectors in the
laboratory-fixed frame, and
there is a constant relative angle $\gamma$ maintained between ${\bf e}_+$
and ${\bf e}_-$.  The averages of the first two products in Eq.
(\ref{eq:e3.7}) are
\cite{zare}
\begin{eqnarray}
\overline{({\bf e}_\pm \cdot {\bf e}_L)^2({\bf e}_\pm \cdot {\bf e}_d)^2}
& = & \frac{1}{5} \; for \; {\bf e}_L \parallel {\bf e}_d \nonumber \\
& = & \frac{1}{15} \; for \; {\bf e}_L \perp {\bf e}_d \: . \label{eq:ea1}
\end{eqnarray}
The average over the crossterm
$({\bf e}_+ \cdot {\bf e}_L)({\bf e}_- \cdot {\bf e}_L)
({\bf e}_+ \cdot {\bf e}_d)({\bf e}_- \cdot {\bf e}_d)$
can be calculated by expressing
${\bf e}_-$ in terms of ${\bf e}_+$ and ${\bf e}_{+\perp}$, a vector
perpendicular to ${\bf e}_+$
\begin{equation}
{\bf e}_- = {\bf e}_+ \cos{\gamma} + {\bf e}_{+\perp} \sin{\gamma} \: .
\label{eq:ea2}
\end{equation}
Substituting Eq. (\ref{eq:ea2}) for ${\bf e}_-$ enables us to express the
above product as a sum of products where the molecule-fixed unit vectors
are either parallel or perpendicular to each other. The average can then
be calculated to give \cite{zare}
\begin{eqnarray}
\overline{({\bf e}_+ \cdot {\bf e}_L)({\bf e}_- \cdot {\bf e}_L)
({\bf e}_+ \cdot {\bf e}_d)({\bf e}_- \cdot {\bf e}_d)} & = &
\frac{1}{5}\cos^2{\gamma} + \frac{1}{15}\sin^2{\gamma} \; for \;
{\bf e}_L \parallel {\bf e}_d \nonumber \\
& = & \frac{1}{15}\cos^2{\gamma} - \frac{1}{30}\sin^2{\gamma} \; for \;
{\bf e}_L \perp {\bf e}_d \; . \label{eq:ea3}
\end{eqnarray}

Eq. (\ref{eq:e3.8}) is obtained by noting that for the case of identical
chromophores,
we have $\mu_1 = \mu_2 \equiv \mu$ and ${\bf e}_+ \perp {\bf e}_-$, (i.e.
$\gamma = \pi/2$). Therefore, using Eq. (\ref{eq:e3.mupmperp}),  we can write
\begin{equation}
\mu_\pm^2 = \mu^2 \langle 0_g0_g|0_e0_g \rangle^2
(1 \pm {\bf e}_1 \cdot {\bf e}_2) \; . \label{eq:ea4}
\end{equation}
Defining $\cos{\phi} \equiv {\bf e}_1 \cdot {\bf e}_2$, and substituting
Eq. (\ref{eq:ea4}) into Eq. (\ref{eq:e3.rsimple}),
Eq. (\ref{eq:e3.8}) is obtained.

The orientational averages required for the evaluation of Eqs. (\ref{eq:e4.13})
and (\ref{eq:e5.8}) can be
similarly evaluated by writing
\begin{equation}
{\bf e}_2 = {\bf e}_1\cos{\phi} + {\bf e}_{1\perp}\sin{\phi} \; ,
\label{eq:ea5}
\end{equation}
which makes it possible to express all orientational averages in terms
of products of averages involving only parallel or perpendicular unit
vectors in the molecule-fixed frame.  The averages needed in
Eqs. (\ref{eq:e4.13})
and (\ref{eq:e5.8}) are
\begin{eqnarray}
\overline{({\bf e}_{1(2)} \cdot {\bf e}_L)^2
({\bf e}_{1(2)} \cdot {\bf e}_d)^2} & = & \frac{1}{5} \; for \;
{\bf e}_L \parallel {\bf e}_d \nonumber \\
& = & \frac{1}{15} \;  for \;
{\bf e}_L \perp {\bf e}_d  \; , \nonumber
\end{eqnarray}
\begin{eqnarray}
\overline{({\bf e}_{2(1)} \cdot {\bf e}_L)^2
({\bf e}_{1(2)} \cdot {\bf e}_d)^2} & = &
\frac{1}{5}\cos^2{\phi} + \frac{1}{15}\sin^2{\phi}  \;  for \;
{\bf e}_L \parallel {\bf e}_d \nonumber \\
& = &
\frac{1}{15}\cos^2{\phi} + \frac{2}{15}\sin^2{\phi}  \;  for \;
{\bf e}_L \perp {\bf e}_d  \; , \nonumber
\end{eqnarray}
\begin{eqnarray}
\overline{({\bf e}_{2(1)} \cdot {\bf e}_L)({\bf e}_{1(2)} \cdot {\bf e}_L)
({\bf e}_{1(2)} \cdot {\bf e}_d)({\bf e}_{2(1)} \cdot {\bf e}_d)}
& = &
\frac{1}{5}\cos^2{\phi} + \frac{1}{15}\sin^2{\phi}  \;  for \;
{\bf e}_L \parallel {\bf e}_d \nonumber \\
& = &
\frac{1}{15}\cos^2{\phi} - \frac{1}{30}\sin^2{\phi}  \;  for \;
{\bf e}_L \perp {\bf e}_d \; ,\nonumber
\end{eqnarray}
and
\begin{eqnarray}
\overline{({\bf e}_{2(1)} \cdot {\bf e}_L)^2
({\bf e}_{1(2)} \cdot {\bf e}_d)({\bf e}_{2(1)} \cdot {\bf e}_d)}
& = &
\frac{1}{5}\cos{\phi}  \;  for \;
{\bf e}_L \parallel {\bf e}_d \nonumber \\
& = &
\frac{1}{15}\cos{\phi}  \;  for \;
{\bf e}_L \perp {\bf e}_d \; ,\label{eq:ea6}
\end{eqnarray}

\newpage

\input fig1

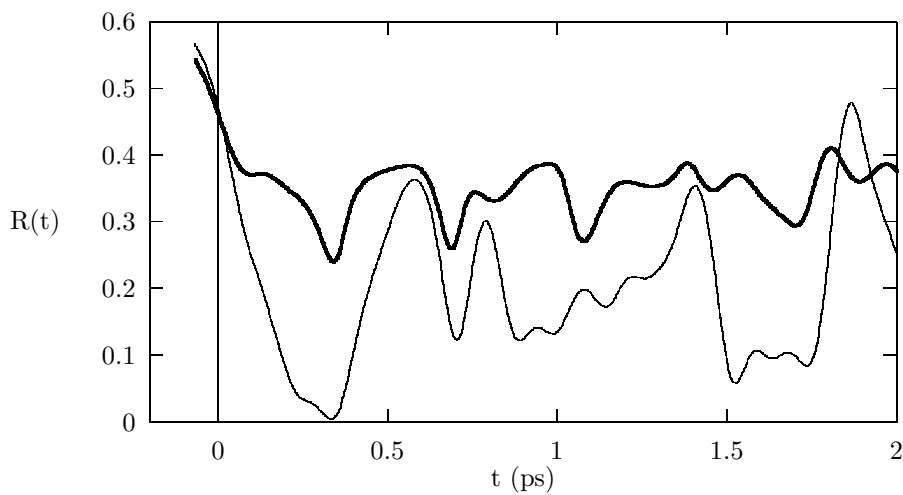
Figure 1. Time-resolved anisotropy plots for two different
excitation/detection conditions.  Upper (thick) curve corresponds to
excitation at zero-zero transition
frequency of lower-energy chromophore, lower (thin) curve is for
excitation 100 cm$^{-1}$ above the zero-zero transition frequency of
higher-energy chromophore.  Detector, with a time window of 70 fs,
is red-shifted from excitation by 250 cm$^{-1}$.

\input fig2

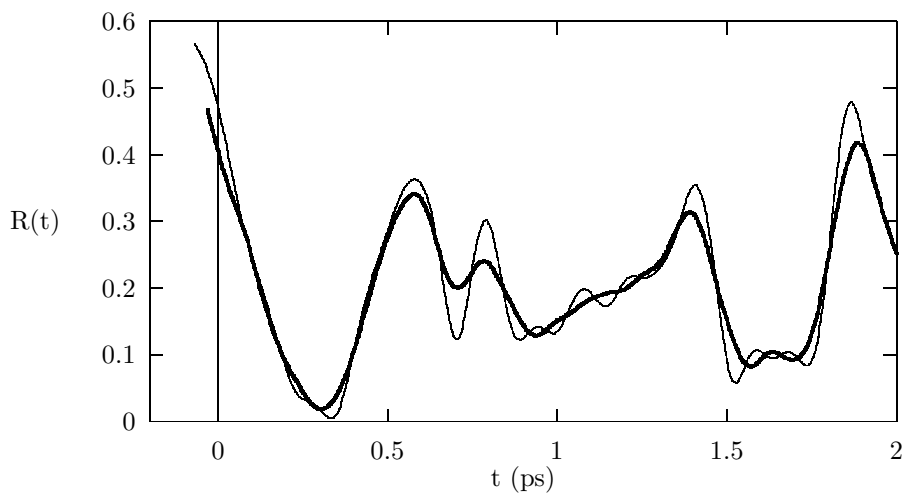
Figure 2. Time-resolved anisotropy corresponding to excitation at
100 cm$^{-1}$ above zero-zero transition frequency of
higher-energy chromophore. Thin curve results from time detection
window of 70 fs (same as in Fig. 1), and thick curve is for a time
detection window of 150 fs.

\input fig3

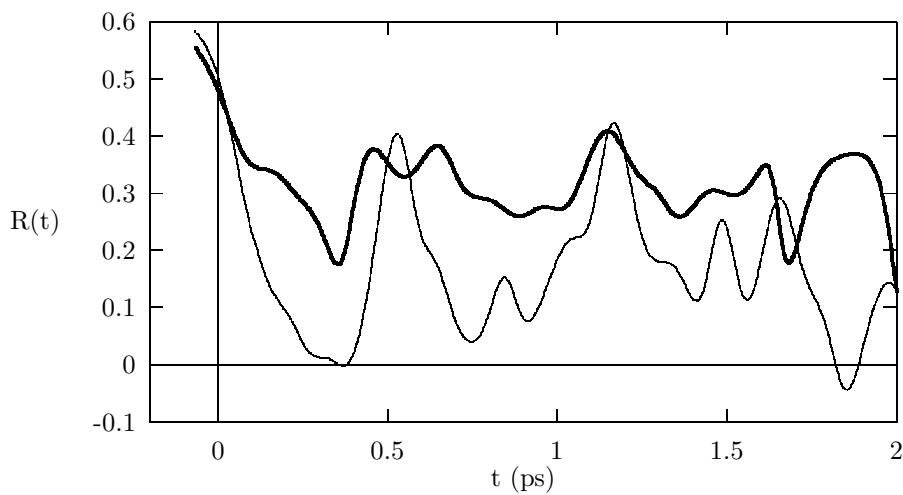
Figure 3. Time-resolved anisotropy for chromophore pair with
100 cm$^{-1}$ difference between the zero-zero
electronic transition frequencies.
Upper (thick) curve corresponds to
excitation at zero-zero transition
frequency of lower-energy chromophore, lower (thin) curve is for
excitation 150 cm$^{-1}$ above the zero-zero transition frequency of
higher-energy chromophore.

\input fig4

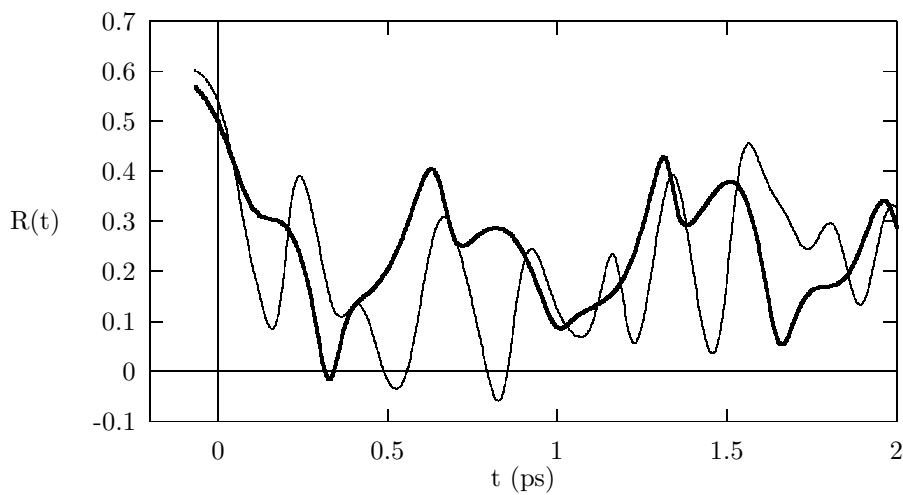
Figure 4. Time-resolved anisotropy for a pair of
identical chromophores.
Upper (thick) curve corresponds to
excitation at zero-zero transition
frequency of the chromophores, lower (thin) curve corresponds to
excitation 250 cm$^{-1}$ above the zero-zero transition frequency.

\input fig5

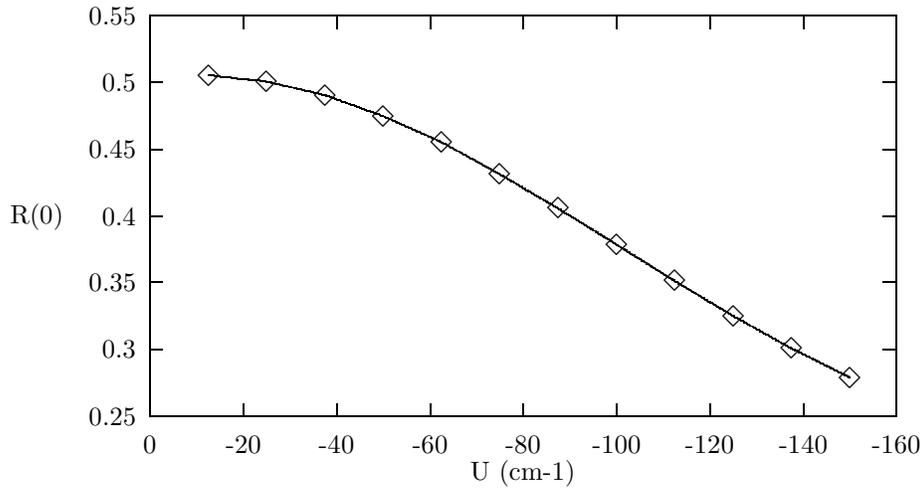
Figure 5. Anisotropy at $t=0$ as a function of excitation
transfer coupling.  Excitation/detection conditions and model system
parameters are the same as in the lower (thin) curve of Figure 1.

\input fig6a

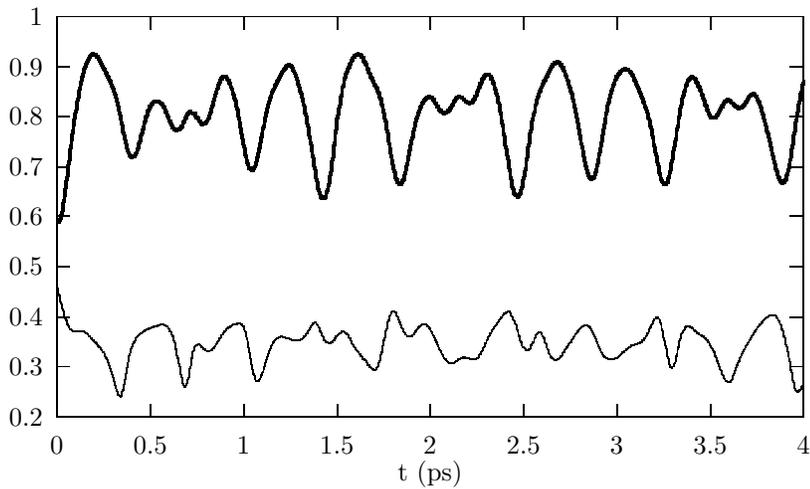
Figure 6a. Thick curve shows the site population difference obtained with
same system parameters and excitation conditions as in the upper (thick)
curve of
Fig. 1.  The thin curve shows the corresponding anisotropy.
Both $\frac{\textstyle P_1(t)-P_2(t)}
{\textstyle P_1(t)+P_2(t)}$ and $R(t)$ are dimensionless.

\input fig6b

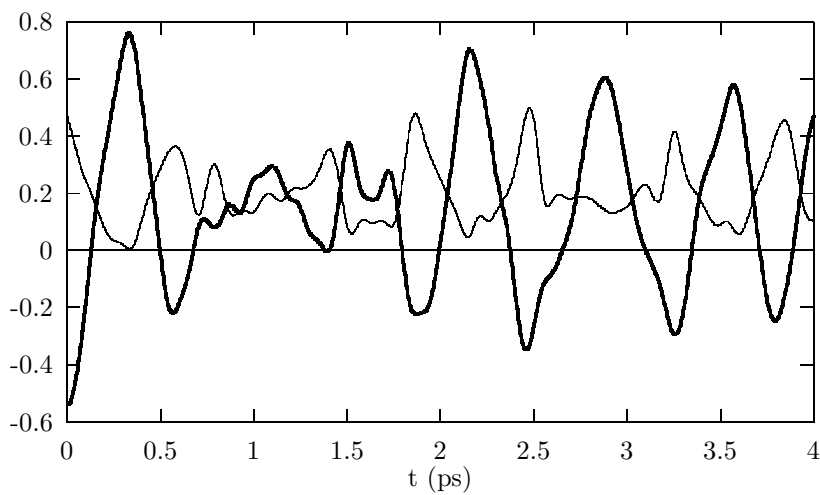
Figure 6b. Thick curve shows the site population difference obtained with
same system parameters and excitation conditions as in the lower (thin)
curve of Fig. 1.
The corresponding anisotropy is shown with a thin curve.

\input fig7

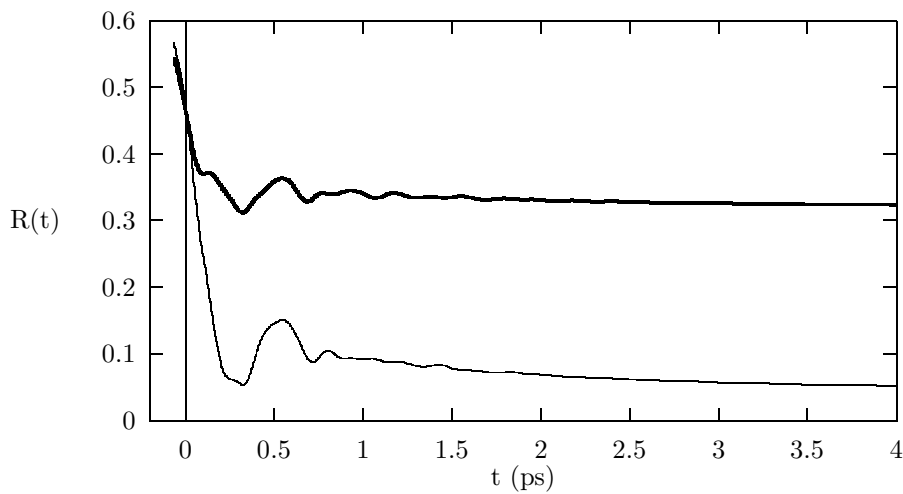
Figure 7. Time-resolved anisotropy for chromophore pair with same
parameters as in Figure 1 and a 1.5 ps$^{-1}$ vibrational relaxation
rate. Upper (thick) curve is for
excitation at zero-zero transition
frequency of lower-energy chromophore, lower (thin) curve is for
excitation 100 cm$^{-1}$ above the zero-zero transition frequency of the higher
energy chromophore. The vibrational and electronic ground state of the
chromophore pair is the only initial state included.

\input fig8

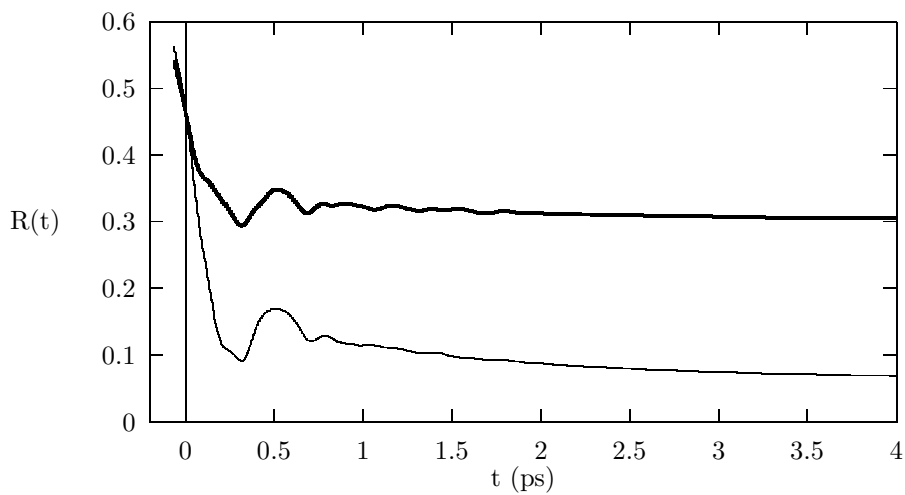
Figure 8.  Time-resolved anisotropy for chromophore pair with the same
parameters as in Figure 1 and a 1.5 ps$^{-1}$ vibrational relaxation
rate.  Three lowest thermally populated initial states have been
included.

\input fig9

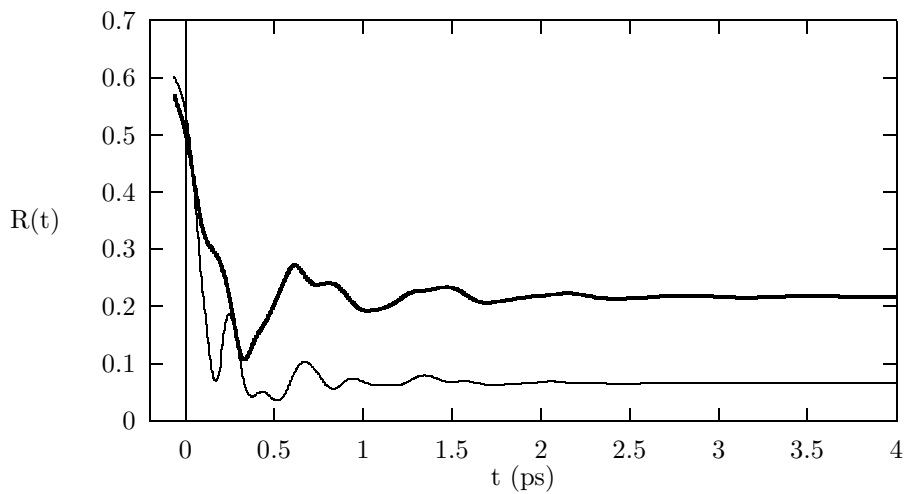
Figure 9. Time-resolved anisotropy for pair of identical chromophores with
the same
parameters as in Figure 4 and a 1.5 ps$^{-1}$ vibrational relaxation
rate. Upper (thick) curve corresponds to
excitation at zero-zero transition
frequency of either chromophore, lower (thin) curve is for
excitation 250 cm$^{-1}$ above the zero-zero transition frequency of either
chromophore.

\input fig10

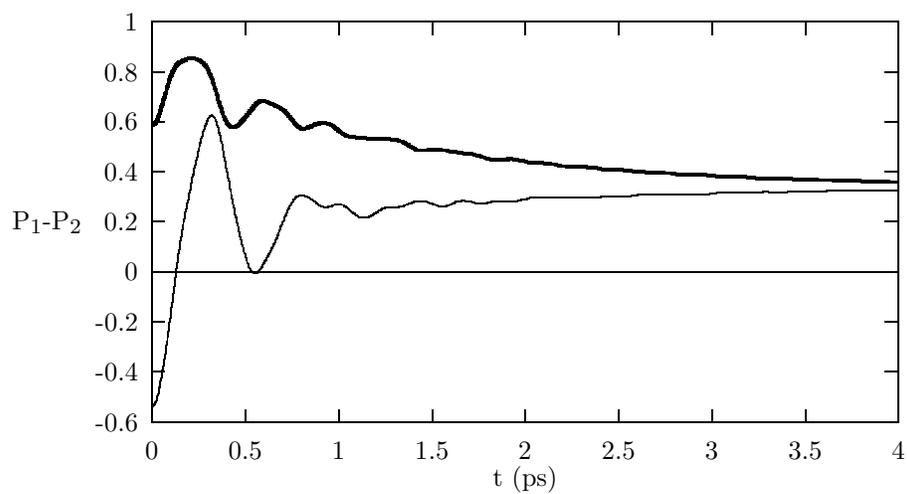
Figure 10. Site population difference for the chromophore pair with
150 cm$^{-1}$ offset between the zero-zero electronic transition frequencies
and a 1.5 ps$^{-1}$ vibrational relaxation
rate.  Upper (thick) curve is for
excitation at zero-zero transition
frequency of lower-energy chromophore, lower (thin) curve corresponds to
excitation 100 cm$^{-1}$ above the zero-zero transition frequency of
higher-energy chromophore.


\begin{references}
\bibitem{sym}
The vibrational modes in our model are assumed to be totally symmetric.
If these modes were not totally symmetric, the vibronic selection rules
would introduce complications in absorption, emission and ultimately, in
the anisotropy.
\end{references}
\end{document}